\documentclass[]{mnras}

\usepackage{graphicx}
\usepackage{color}
\title[The OmegaWhite Survey: II ]{The OmegaWhite Survey for short
  period variable stars II: An overview of results from the first
  four years}

\author[] {R.~Toma$^{1}$\thanks{Email: rto@arm.ac.uk},
  G. Ramsay$^{1}$, S.~Macfarlane$^{2,3}$, P. J~Groot$^2$, P. A.~Woudt$^3$, V. Dhillon$^{4,6}$, \and C. S. Jeffery$^1$,  
T. Marsh$^5$, G. Nelemans$^2$, D. Steeghs$^5$\\ 
$^{1}$Armagh Observatory, College Hill, Armagh, BT61 9DG, UK\\ 
$^2$Department of Astrophysics/IMAPP, Radboud University, P.O. Box 9010, 6500 GL Nijmegen, 
The Netherlands\\ 
$^3$Department of Astronomy, University of Cape Town, Private Bag X3, Rondebosch 7700, South Africa\\ 
$^4$Department of Physics \& Astronomy, University of Sheffield, Sheffield S3 7RH, UK\\
$^5$Department of Physics, University of Warwick, Coventry CV4 7AL, UK\\
$^6$Instituto de Astrofisica de Canarias, E-38205 La Laguna, Tenerife, Spain\\
}

\date{Accepted 2016 August 16. Received 2016 August 16; in original form 2016 July 15}

\begin{document}
\outer\def\gtae {$\buildrel {\lower3pt\hbox{$>$}} \over 
{\lower2pt\hbox{$\sim$}} $}
\outer\def\ltae {$\buildrel {\lower3pt\hbox{$<$}} \over 
{\lower2pt\hbox{$\sim$}} $}
\newcommand{\Msun} {$M_{\odot}$}
\newcommand{\Rsun} {$R_{\odot}$}
\newcommand{\solar} {${\odot}$}
\newcommand{\kep}{\it Kepler}
\newcommand{\swift}{\it Swift}
\newcommand{\Porb}{P_{\rm orb}}
\newcommand{\nuorb}{\nu_{\rm orb}}
\newcommand{\eplus}{\epsilon_+}
\newcommand{\eminus}{\epsilon_-}
\newcommand{\cd}{{\rm\ c\ d^{-1}}}
\newcommand{\MdotL}{\dot M_{\rm L1}}
\newcommand{\Ldisk}{L_{\rm disk}}
\newcommand{\src}{KIC 9202990}
\newcommand{\ergscm} {erg s$^{-1}$ cm$^{-2}$}
\newcommand{\rchi}{$\chi^{2}_{\nu}$}
\newcommand{\chisq}{$\chi^{2}$}
\newcommand{\pcmsq} {cm$^{-2}$}
\newcommand{\delSct} {$\delta$ Sct}

\maketitle
\begin{abstract}
OmegaWhite is a wide-field, high cadence, synoptic survey targeting
fields in the southern Galactic plane, with the aim of discovering
short period variable stars. Our strategy is to take a series of 39 s
exposures in the $g$ band of a 1 square degree of sky lasting 2 h
using the OmegaCAM wide field imager on the VLT Survey Telescope
(VST). We give an overview of the initial 4 years of data which covers
134 square degrees and includes 12.3 million light curves.  As the
fields overlap with the VLT Survey Telescope H$\alpha$ Photometric
Survey of the Galactic plane and Bulge (VPHAS+), we currently have
$ugriH\alpha$ photometry for $\sim$1/3 of our fields. We find that a
significant fraction of the light curves have been affected by the
diffraction spikes of bright stars sweeping across stars within a
  few dozen of pixels over the two hour observing time interval due
to the alt-az nature of the VST. We select candidate variable stars
using a variety of variability statistics, followed by a manual
verification stage. We present samples of several classes of short
period variables, including: an ultra compact binary, a DQ white
dwarf, a compact object with evidence of a 100 min rotation period,
three CVs, one eclipsing binary with an 85 min period, a symbiotic
binary which shows evidence of a 31 min photometric period, and a
large sample of candidate {\delSct} type stars including one with a
9.3 min period. Our overall goal is to cover 400 square degrees, and
this study indicates we will find many more interesting short period
variable stars as a result.
\end{abstract}


\section{Introduction}

The aim of the OmegaWhite (OW) survey is to discover a population of
short period variable stars close to the Galactic plane (Macfarlane et
al. 2015, Paper I). In particular, our main goal is to discover ultra
compact binaries (UCBs) which have orbital periods shorter than 70
min. UCBs can take various guises including: X-ray binaries containing
white dwarf -- neutron star components (see e.g. Nelemans \& Jonker
2010), sub-dwarf B star -- white dwarf components (e.g. Geier et
al. 2013), double degenerate non-interacting binaries (e.g. Brown et
al. 2011, 2016, Gianninas et al. 2015), and the double degenerate
interacting binaries (the AM CVn stars, see Solheim 2010 for a
review).

\begin{table*}
\centering
\caption{The log for OW observations made during ESO
Semesters 88--94. We show the calendar date, the number of fields
which were observed and the range in their sky co-ordinates, and
the number of light curves (prior to the flagging stage) which
were obtained in each Semester.}
\begin{tabular}{cccccc}
\hline
ESO  & Date & Observed & RA & DEC & Number of \\
Period  &  & Fields & (J2000) & (J2000) & light curves  \\
 &  & & (hh:mm) & (\degr:\arcmin) & ($\times10^{6}$) \\
\hline
88   & Dec 2011 $\rightarrow$ Apr 2012 & 26 & 07h35m $\rightarrow$ 08h25m & --30\degr00\arcmin $\rightarrow$ --26\degr00\arcmin & 1.7 \\
90   & Nov 2012 $\rightarrow$ Mar 2013 & 4 & 07h05m $\rightarrow$ 08h25m  & --30\degr00\arcmin $\rightarrow$ --25\degr00\arcmin & 0.2  \\
91   & Apr 2013 $\rightarrow$ Sep 2013 & 26 & 17h00m $\rightarrow$ 18h25m  & --29\degr30\arcmin $\rightarrow$ --23\degr30\arcmin  & 3.5\\
92   & Dec 2013 $\rightarrow$ Apr 2014 & 8 & 07h05m $\rightarrow$ 08h40m  & --31\degr00\arcmin $\rightarrow$ --26\degr00\arcmin  & 0.3\\
93 & Apr 2014  $\rightarrow$ Sep 2014 &  34 & 17h05m $\rightarrow$ 18h30m  & --32\degr30\arcmin $\rightarrow$ --21\degr30\arcmin  & 4.5 \\
94 & Dec 2015  $\rightarrow$ Apr 2015 &  36 & 07h15m $\rightarrow$ 08h30m & --32\degr00\arcmin $\rightarrow$ --23\degr00\arcmin & 2.1\\
\hline
\end{tabular}
\label{obslog}
\end{table*}

UCBs are predicted to be strong persistent sources of low-frequency
gravitational waves accessible with facilities such as the 
Evolved Laser Interferometer Space Antenna (eLISA, e.g. Nelemans
2013). Early estimates of the intrinsic number of AM CVn binaries
suggested a relatively high gravitational wave background. However, in
a series of papers using Sloan Digital Sky Survey (SDSS) data
(York et al. 2000) it was established that the observed space density
of AM CVn stars is 5$\pm3\times 10^{-7} $pc$^{-3}$ (Carter et
al. 2013): a factor of 12 lower than the `pessimistic' model
predictions outlined in Roelofs, Nelemans \& Groot 2007 which were
based on the models of Nelemans et al. (2001, 2004). The SDSS work led
to the discovery of AM CVn stars with orbital periods in the range 35
min $<P_{orb}<$ 65 min (e.g. Roelofs et al. 2005, Anderson et
al. 2005, 2008, Carter et al. 2014a, 2014b). More recently, the
Palomar Transient Factory (PTF) survey has identified outbursting AM
CVn binaries in the 20-35 minute period range (Levitan et al. 2011,
2013, 2014). It is systems with the shortest orbital period ($<$ 20
min) which are predicted to be the strongest emitters of gravitational
waves. Determining their number is important for the development of
eLISA and for understanding the relative importance of the three
predicted formation channels of these binaries. One way to identify
those AM CVn stars with the shortest orbital period is through their
photometric behaviour as they often show a modulation on a period at,
or close to, the orbital period (e.g. HM Cnc, Roelofs et al. 2010; AM
CVn itself, Nelemans, Steeghs \& Groot 2001; SDSS J190817+3940, Kupfer
et al. 2015).

\begin{figure*}
\begin{center}
\setlength{\unitlength}{1cm}
\begin{picture}(12,6)
\put(-3.4,0){\includegraphics{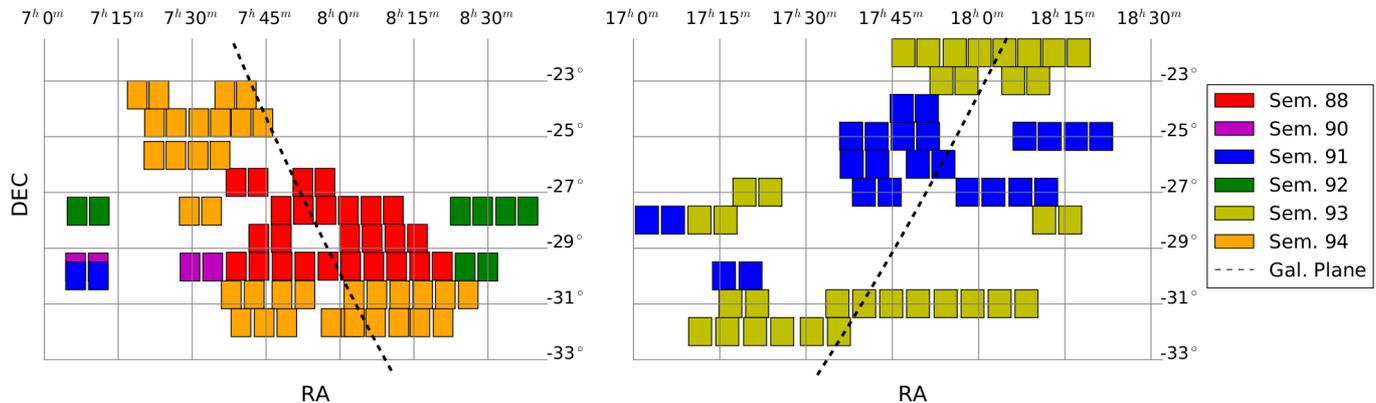}}
\end{picture}
\end{center}
\caption{The two regions of OW field pointings along the Galactic
  plane (dashed line) and the Galactic Bulge region (right). We have
  colour coded fields to indicate in which Semester the data were
  obtained.}
\label{pointings} 
\end{figure*}

The OW survey has a similar strategy to the RApid Temporal Survey
  (RATS, Ramsay \& Hakala 2005, Barclay et al. 2011, Ramsay et al.
2014). However, in contrast to the initial RATS survey, which had a
sky coverage of 40 square degrees, OW has a goal of 400 square degree
coverage. With a typical cadence of 3.6 min, OW has a much higher
cadence than the vast majority of wide field synoptic surveys. For
instance, the PTF typically has a 5 day cadence (Law et al. 2009), and
in the southern hemisphere, {\it Skymapper} (Keller et al. 2007) has a
cadence of hours to years. {\sl Kepler/K2} is able to provide 1 min
cadence but only for a small number of targets at any time (Gilliland
et al. 2010).  Additionally, unlike OW, most surveys exclude the
Galactic Plane because of the high stellar density.

High cadence observations of fields along the Galactic plane will also
deliver a comprehensive census of short period variable stars in the
Galactic plane. These range from pulsating white dwarfs, sdB stars,
$\delta$ Sct stars and their cousins the SX Phe and $\gamma$ Dor
stars, as well as longer period systems such as contact binaries and
transient phenomena such as flare stars.

In this paper, we outline the initial results of the first 4 years
(2011 December to 2015 April) of the OW survey. We briefly summarise
the reduction strategy, but highlight specific issues such as the
process of flagging light curves which are likely to show spurious
variability. We show some examples of new variables which are either
examples of rare variable stars or can illuminate details of physical
processes at work. In a companion paper (Macfarlane et al. 2016, Paper
III) we present followup photometric and spectroscopic observations of
variable stars identified in the OW survey which were made at the
South African Astronomical Observatory, Sutherland, South Africa.

\section{Observations}

\subsection{Overview}

Observations are made using the 2.65-m VLT Survey Telescope (VST, 
Capaccioli \& Schipani 2011, Schipani et al. 2012) located
at ESO's Paranal Observatory in Chile, and the OmegaCAM mosaic imager
instrument (Kuijken 2011, Mieske et al. 2013) which has a field of
view of one square degree which is covered by 32 CCDs with a pixel
scale of 0.22 arcsec/pixel. We take a sequence consisting of one 39 s
exposure; move the telescope to an adjacent field; take another 39 s
exposure; return to the first field and repeat for 2 h in total (all
images are taken in the $g$ band). The mean cadence of images
  obtained in the Semesters covered in this paper is 3.6$\pm$0.7 min,
  with a mean minimum cadence of 2.8 min, where variations in the mean
  cadence are likely due to differences in the time it takes to do
  focussing and mirror alignment over the 2 h observation. In Paper I
we presented the results of data obtained during Semester 88. In this
paper we present an overview of the data obtained during Semesters 88
and 90--94. For reference, we show in Table \ref{obslog} the number of
fields observed and in Figure \ref{pointings} the sky position of
these fields. In Table \ref{fieldcentres} we give details of the
individual fields which were observed during Semester 90--94 (Paper I
outlines the fields observed in Semester 88). Combined, these data
cover 134 square degrees of sky and we obtain 12.3$\times10^{6}$ light
curves.

\begin{figure*}
\begin{center}
\setlength{\unitlength}{1cm}
\begin{picture}(12,6)
\put(-3,0){\includegraphics{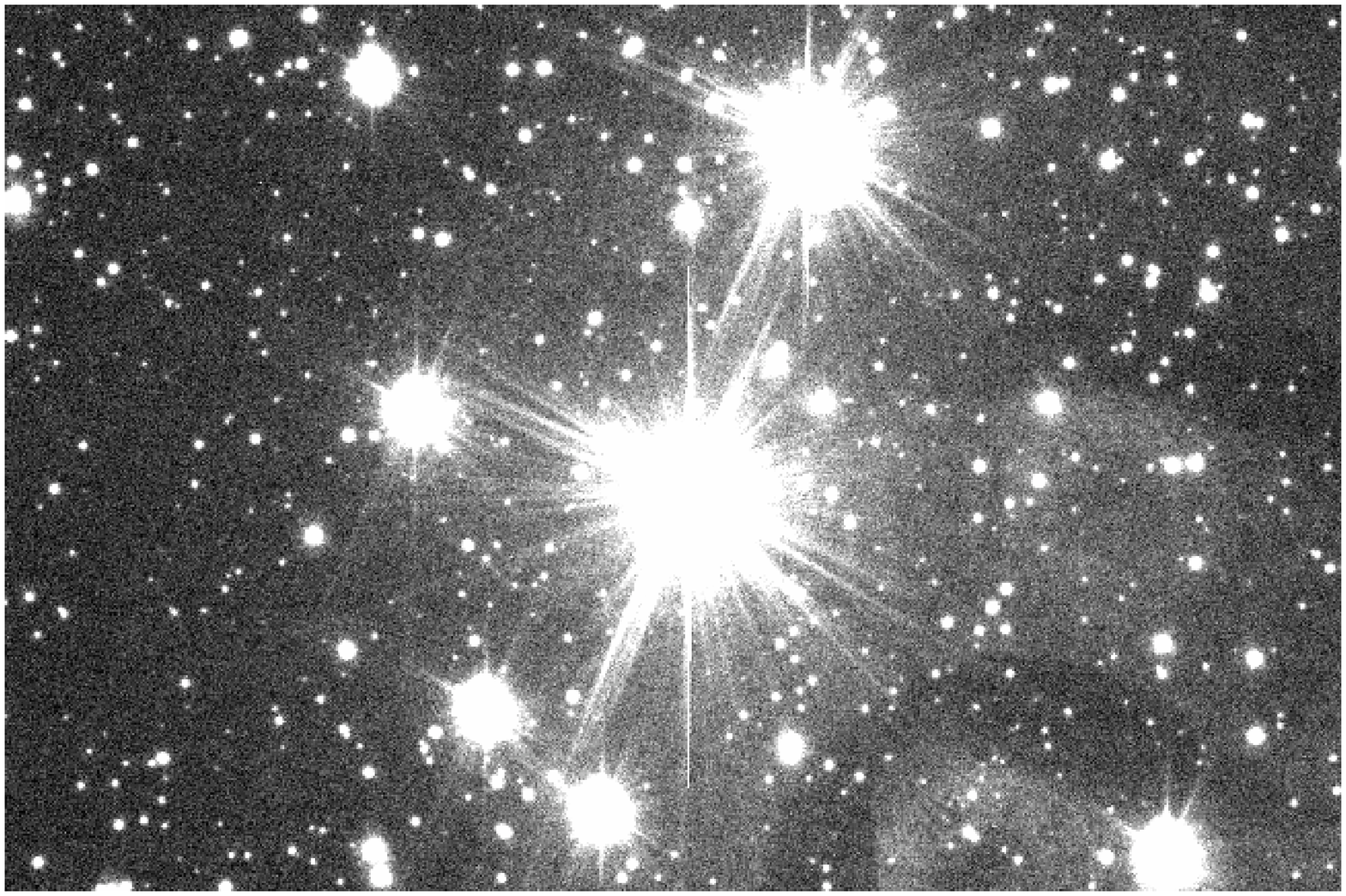}}
\put(6,0){\includegraphics{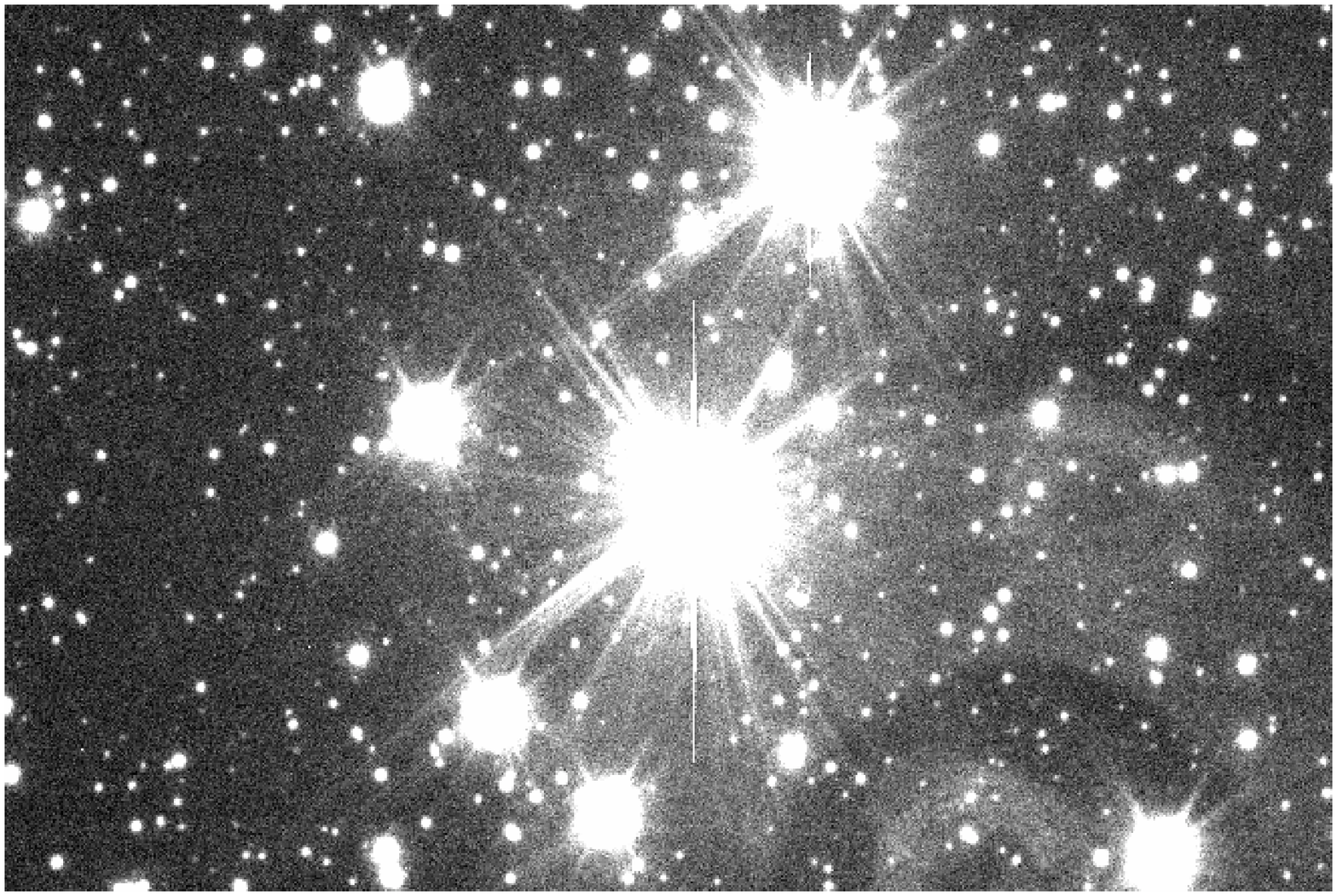}}
\end{picture}
\end{center}
\caption{Two images from Field1a, Chip71 taken in Semester 88, (center
  of the image is $\alpha_{J2000.0}$=07:40:26, $\delta_{J2000.0}$=--30:04:18)
  where the rotation of the diffraction spikes relative to the
  detector is clear. The effect of these spikes sweeping across stars
  can result in spurious variability.}
\label{spikes} 
\end{figure*}

\subsection{Data Reduction}

Paper I presents details of our data reduction process which includes
the image cleaning process, the photometric reduction, how we flag
light curves which may have been contaminated by instrumental effects,
and how we identify objects which are variable. For brevity, we only
give a summary of these steps and the reader is directed to Paper I
for more details of the individual stages. We highlight issues which
may cause a degradation of the light curves.

\begin{enumerate}

\item Data are downloaded from the ESO archive and the images are
  cleaned using a set of master bias images which have been made on
  the same day as the science images and a set of master flat fields
  which were made from observations obtained during the month of the
  science observations. An astrometic solution is obtained for each
  image using {\tt ASTROMETRY.NET} software (Lang et al. 2010). The
  typical residual to the 2MASS positions is $\sim$0.1 arcsec (see
  Paper I for details).

\item Since our fields lie within $\sim$5 degrees of the Galactic
  Plane, many (but not all) of the fields are crowded. To account for
  this and to accommodate variations in seeing, we use the Difference
  Image Analysis Package {\tt DIAPL2} (Wozniak 2000) to obtain our
  photometry. Most wide field photometric surveys show systematic
  trends in the light curves (i.e. they all show the same long term
  features) which are due to variations in seeing, transparency and
  airmass. We detrend our light curves using the {\tt SYSREM}
  algorithm (Tamuz, Mazeh \& Zucker 2005). We place the magnitude of
  each star (OW$g$) on to the standard (Vega) system by
  cross-calibrating the reference image for each field with the
    AAVSO Photometric All-Sky Survey (APASS) $g$ band all-sky
  catalogue (Henden et al. 2012).

\item Paper I described a method to flag stars which have been falsely
  identified as variable stars. The source of these false positives
  include proximity to the detector edge, bad pixels or saturated
  stars, or have local high background. The origin of the latter
  effect can be due to diffraction spikes from saturated stars. Since
  the VST has an altitude-azimuth mounting, these spikes move over the
  2 h time interval that the observations are made (see Figure
  \ref{spikes} for an example). This can inject a spurious modulation
  into light curves.  There is also some movement of stars over the
  detector (we do not auto-guide the telescope), which can cause stars
  to cross areas of bad or hot pixels.

\item Photometric variability parameters are determined for all light
  curves using the {\tt VARTOOLS} package (Hartman et al. 2008).
  These parameters include: the RMS, $\chi^{2}_{\nu}$, the period (P$_{LS}$)
  corresponding to the highest peak in the Lomb Scargle
  (LS) Power spectrum (Lomb 1976, Scargle 1982, Zechmeister \&
    K\"{u}rster 2009, Press et al. 1992), and its associated False
  Alarm Probability (FAP), and the period and associated FAP from the
  Analysis of Variance (AoV) period search algorithm (Schwarzenberg-Czerny
    1989, Devor 2005) which identifies a period by folding the data on
    a number of trial periods and splits the folded data into
    different phase bins. The FAP gives a measure of how likely a
  peak in a periodogram is due to random noise in a light curve. For
  the LS analysis we searched in frequency space between the Nyquist
  frequency and the frequency corresponding to the duration of the
  light curve (these parameters varied by a small degree from
  field-to-field).  Variable objects are then selected on a field by
  field basis using the distribution of stars in the P$_{LS}$, FAP
  plane. We outline how we identify variable stars in the next
  section.

\end{enumerate}

\section{Identifying variable stars}

How to best identify variable stars among a sample of stars has been
explored by many authors (see e.g. Graham et al. 2013). Specific
variability tests are best suited to find particular types of variable
star. Perhaps the most straightforward method is to select them
according to the RMS of their light curve as a function of magnitude
-- variable objects will have a higher RMS compared to non-variable
objects with a similar magnitude.  For stars which show periodic flux
variations, the LS periodogram is efficient in identifying them. Since
we are aiming to find periodic variables we primarily use the LS test.

\begin{table*}
\centering
\caption{The number of variable candidates selected using the Lomb
  Scargle periodogram in each sample defined by four MAD '$n$' levels.
  The candidates have been grouped in period ranges appropriate for
  different broad classes of variable star. We indicate the number of
  stars which have been flagged as having a high background which is a
  likely indicator that a bright star is nearby and may have caused
  the light curve to be affected by rotating diffraction spikes (see
  Figure 2).}
 \begin{tabular}[pos]{rrrrrrrr}  
\hline
$n$ & 0--20 & 20--40 & 40--60 & 60--110 & $>$110 & Total & Stars with  \\
    & (min)  & (min) & (min)  & (min) & (min) & (stars) & High Background\\
\hline
5 & 1673 & 1495 & 5210 & 20462 & 2553 & 31389 & 5129 (16.3\%)\\
10 & 1643 & 1355 & 3318 & 7383 & 518 & 14213 & 2658 (18.7\%) \\
15 & 348 & 473 & 983 & 1870 & 112 & 3782 & 613 (16.2\%) \\
20 & 14  & 137 & 248 & 482 & 32 & 909 & 140 (15.4\%) \\
\hline
 \end{tabular}
\label{MADsamples}
\end{table*}

Our primary method for identifying periodic variable stars relies on
the distribution of stars in the P$_{LS}$, FAP plane. We note that
large samples will have a correspondingly large number of false
positives purely from statistical reasons. For instance, there are
4.1$\times10^{6}$ light curves which have P$_{LS}<20$ min. If we were
to apply a threshold of log$_{10}$(FAP)$<$--2.5 (i.e. a probability of
0.316 percent of being a false positive), we would expect 13110 false
positives (out of 4.1$\times10^{6}$ light curves). Rather than using a
fixed threshold of log$_{10}$(FAP) to provide an initial selection of
candidate variables, we used the Median Absolute Deviation (MAD) to
provide a means of identifying stars which are `outliers' in the
P$_{LS}$, FAP plane (c.f. Figure 3). The MAD statistic is defined for
a batch of parameters $\{x_{1},\ldots,x_{m}\}$ as
\begin{equation}
 \mathrm{MAD} = \mathrm{median}( | x_{i} - \mathrm{median} (x_{i}) | )
\end{equation}
where $x_{i}$, for our purposes, represents values of the FAP (each
with a corresponding value of P$_{LS}$). The data are sorted according
to P$_{LS}$ and put into 2 min period bins and the MAD of the FAP in
each bin determined. We use a multiplier, $n$, to select samples which
are progressively more distant from the local median FAP of points in
each time bin. We do not have an {\it a priori} knowledge of which
value of $n$ is the most appropriate to use. We therefore identify
four samples with n=5, 10, 15, and 20 which obey log$_{10}(FAP)<$
Median(log$_{10}$FAP)--(MAD(log$_{10}$FAP)$\times{n}$). This test is
done on a field by field basis. For brevity we use the term ``MAD$n$
sample'' to indicate which value of $n$ was used.

\begin{figure*}
\begin{center}
\setlength{\unitlength}{1cm}
\begin{picture}(12,12.)
\put(-2.8,6.2){\includegraphics{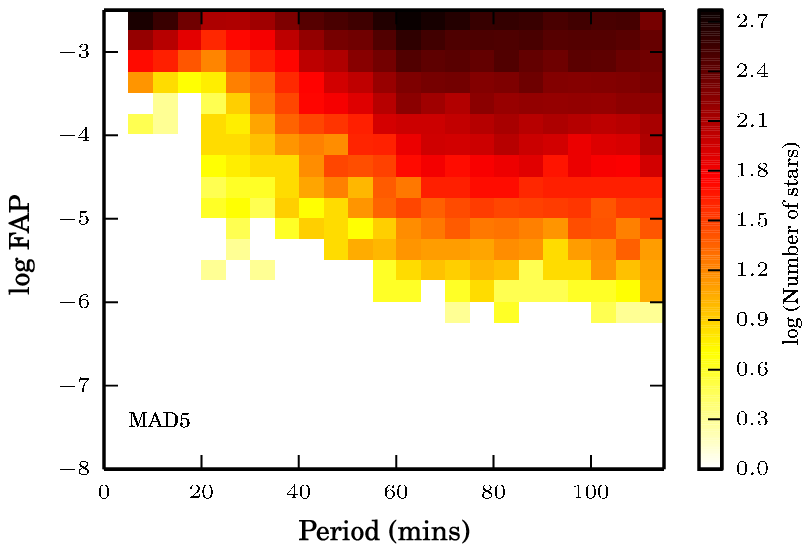}}
\put(5.8,6.2){\includegraphics{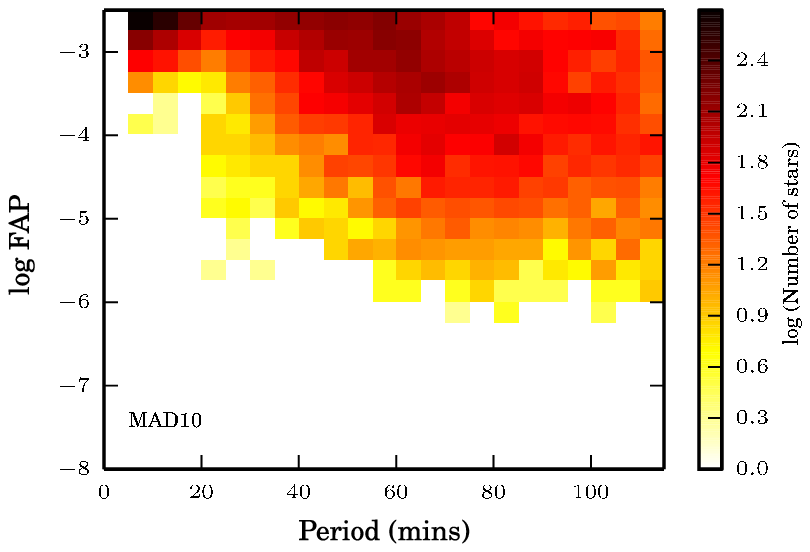}}
\put(-2.8,-0.3){\includegraphics{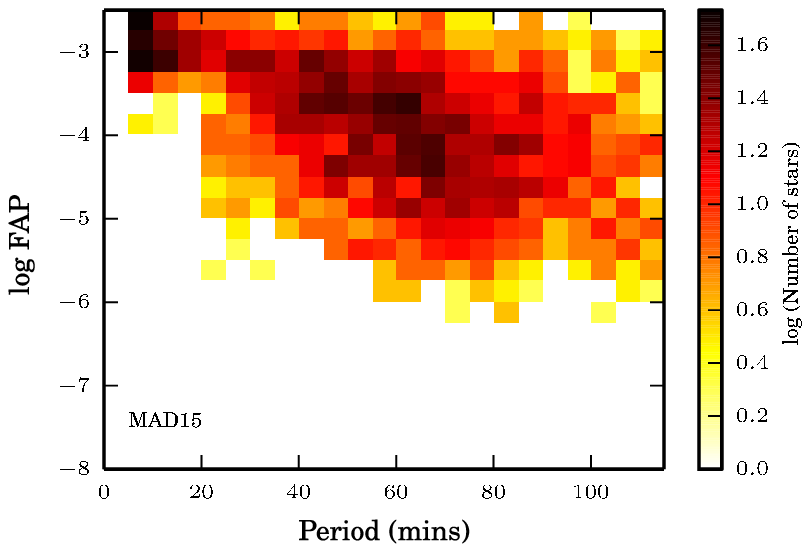}}
\put(5.8,-0.3){\includegraphics{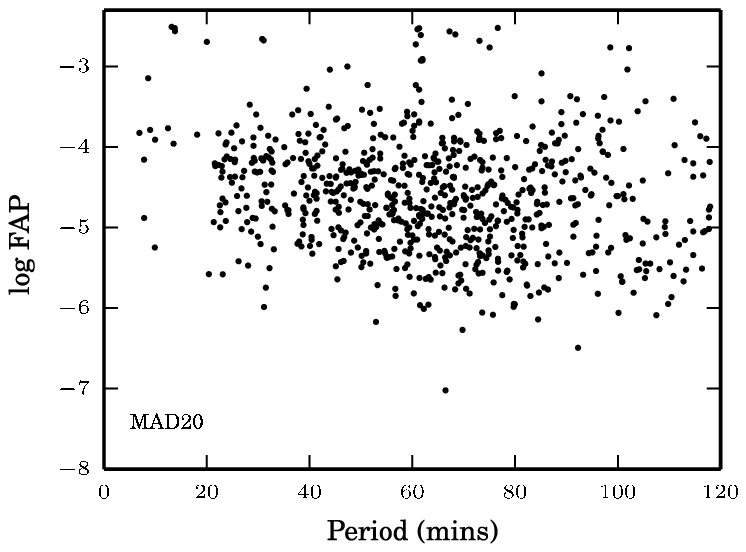}}
\end{picture}
\end{center}
\caption{The distribution of variable candidates from Semester 88,
  90-94 as function of log$_{10}$FAP and period for our four samples
  (stars which have been automatically flagged as having spurious
  variability have been removed). The stars are grouped in bins of 5
  minute $\times$ 0.25 log$_{10}$FAP units. The darker the area, the
  greater the number of stars it contains. The trends of stars are
  discussed in the text. Individual stars are shown in the bottom
  right sample for MAD20. A small number of stars appear with
  relatively high values of log FAP due to the fact that stars are
  identified over a 2 min period bin which may be poorly populated.}
\label{madplots} 
\end{figure*}

As can be seen in Figure \ref{madplots}, which shows the distribution
of the MAD5, MAD10, MAD15 and MAD20 samples, larger values of $n$
gives fewer stars -- i.e. they are variables with the highest
probability. We show the number of stars in different period ranges as
a function of different values of $n$ in Table \ref{MADsamples}.  As
can also be seen from Figure \ref{madplots} the distribution of all
but the MAD20 sample is highly non-uniform.  There are several reasons
for this. Firstly, there are intrinsically more variables with long
periods compared to short periods (for instance $\delta$ Sct variables
which have periods typically greater than half an hour are one of the
most populous of all variable stars).  Secondly, if systematic trends
have been imperfectly removed, longer period trends may remain in the
light curves. There are a significant number of stars with P$_{LS}<$20
min -- these are the stars we are most interested in.

\begin{figure*}
\begin{center}
\setlength{\unitlength}{1cm}
\begin{picture}(12,12)
\put(-2.8,6){\includegraphics{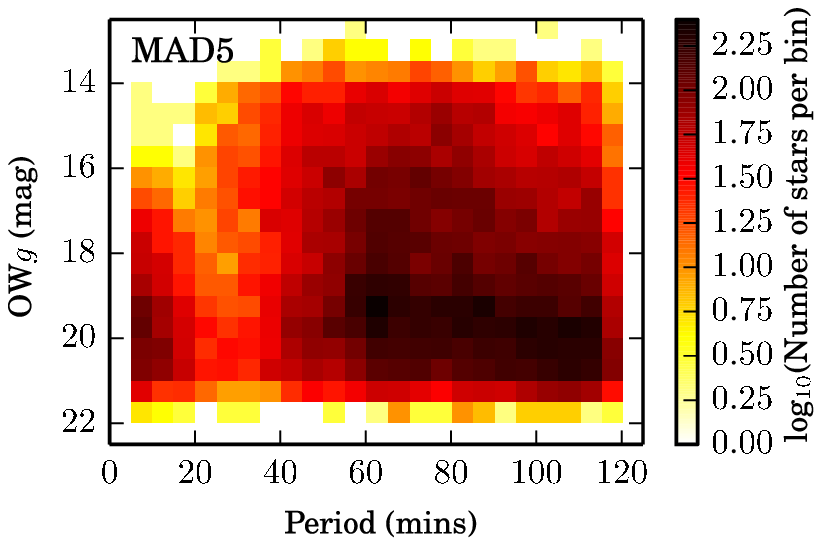}}
\put(5.8,6){\includegraphics{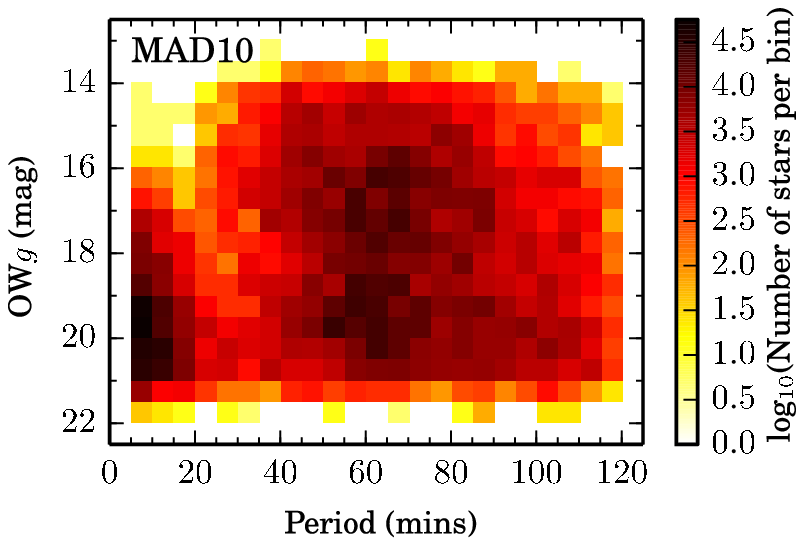}}
\put(-2.8,0){\includegraphics{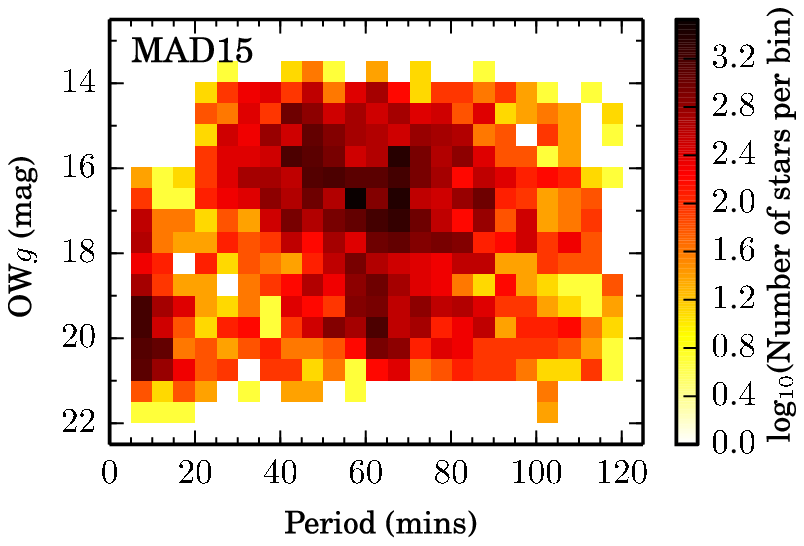}}
\put(5.8,0){\includegraphics{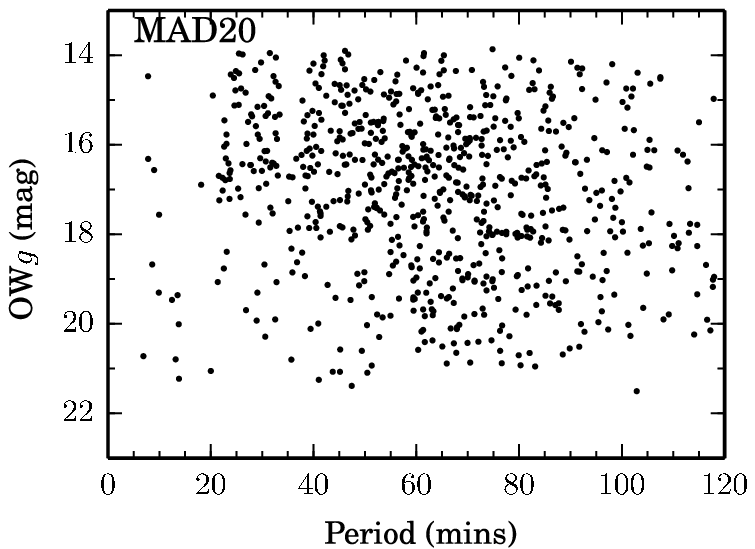}}
\end{picture}
\end{center}
\caption{We show the distribution of variable candidates in the
  Period vs OW$g$ plane for the four samples using the same
  procedure as in Figure~\ref{madplots} (stars which have been
  automatically flagged as having spurious variability have been
  removed). The stars are grouped in bins of 5 minute $\times$ 0.5
  mag. The darker the area, the larger the number of stars it
  contains. The trends of faint variable stars are explained in the
  text.}
\label{magper} 
\end{figure*}

Figure \ref{magper} shows the distribution of the four samples in the
period, OW$g$ plane. Compared to the MAD5 sample, the MAD15 sample has
fewer stars which are faint and have long periods. We expect that this
is due to many relatively faint objects in the MAD5 sample which have
light curves which contain residual systematic trends, or have been
affected by diffraction spikes from bright stars. In contrast, the
MAD15 sample shows an enhancement of relatively bright stars with
periods in the range of 40--70 min: this is likely due to stellar
pulsators. At the short period end of the distribution there is a
concentration of faint stars which is where we expect UCBs to be
present.

\begin{figure*}
\begin{center}
\setlength{\unitlength}{1cm}
\begin{picture}(12,10)
\put(-4.5,-8){\includegraphics{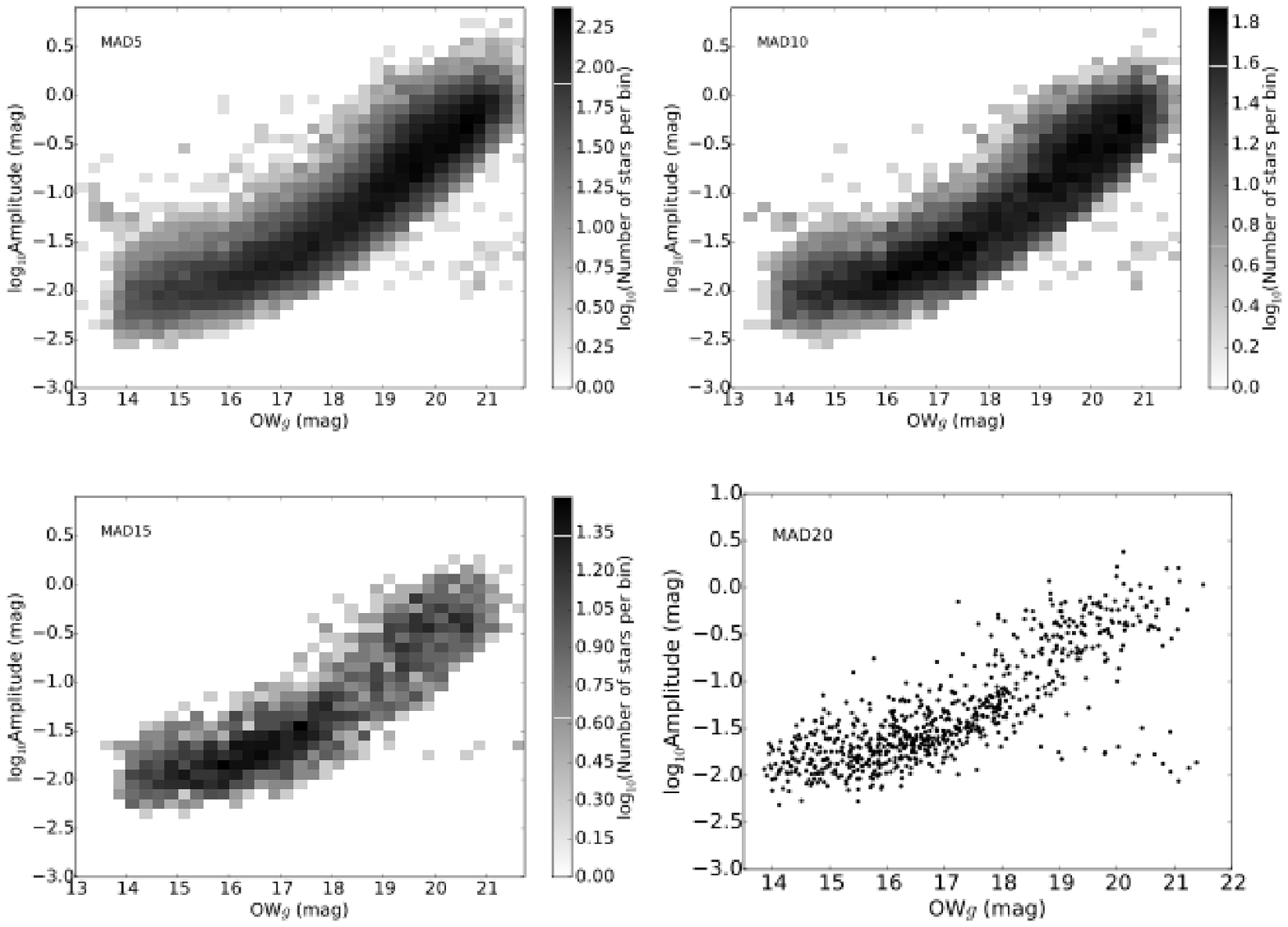}}
\end{picture}
\end{center}
\caption{We show the distribution of variable candidates in the OW$g$
  vs Amplitude plane for the four samples using the same procedure as
  in Figure~\ref{madplots} (stars which have been automatically
  flagged as having spurious variability have been removed). The stars
  are grouped in bins of 0.25 mag $\times$ 0.1 log$_{10}$Amplitude
  units. The darker the area, the larger the number of stars it
  contains. The trends of faint variable stars are explained in the
  text.}
\label{amppermag} 
\end{figure*}

Figure \ref{amppermag} shows the amplitude of the periodic modulation
as a function of magnitude for the four samples. We find that for
fainter stars the amplitude of intrinsic variability needs to be
progressively higher to beat the noise and be selected. The MAD15
sample shows a concentration of low amplitude and relatively bright
stars which is likely due to {\delSct} stars.

\begin{figure}
\begin{center}
\setlength{\unitlength}{1cm}
\begin{picture}(6,6.2)
\put(9.4,-0.7){\includegraphics{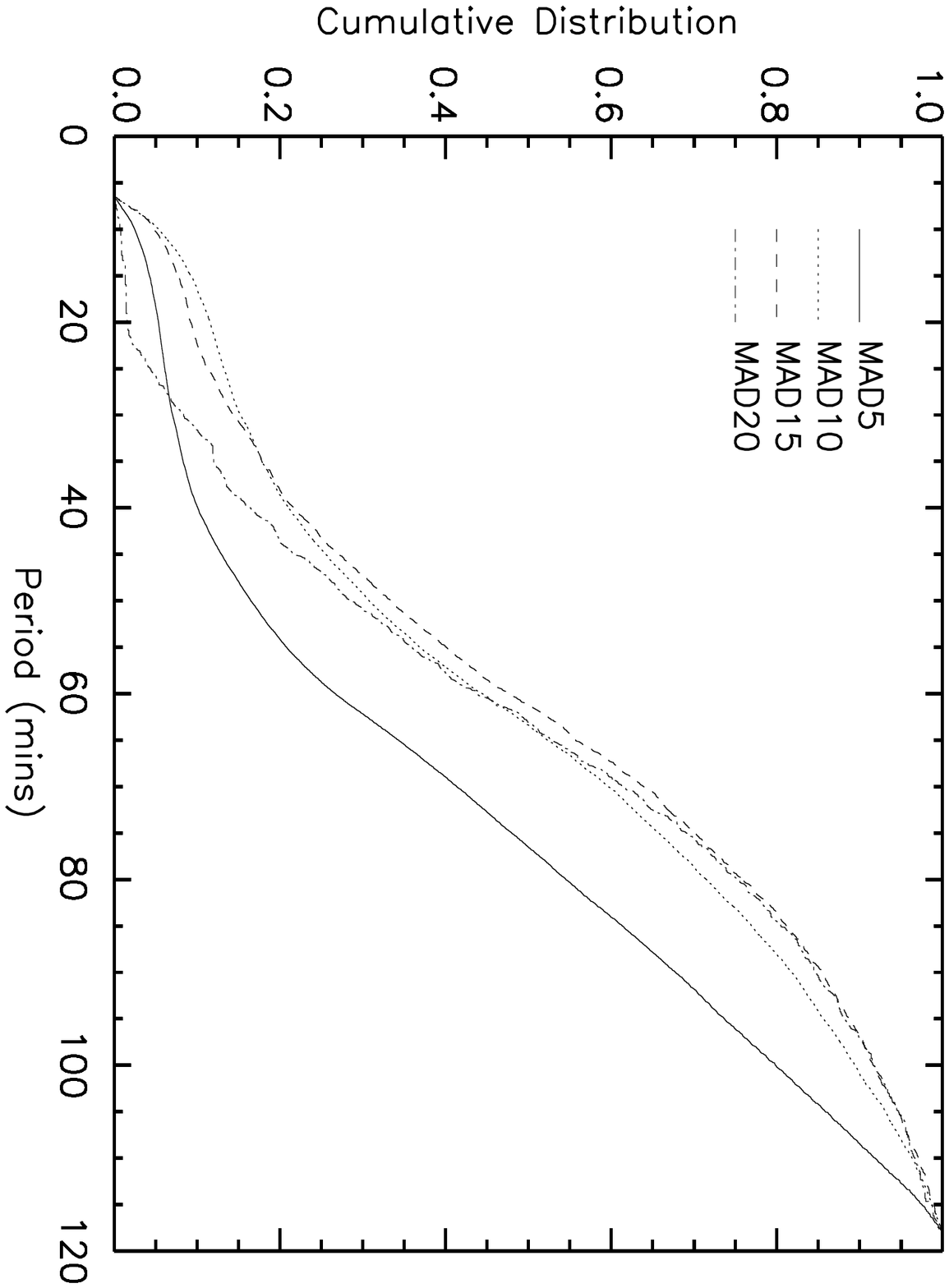}}
\end{picture}
\end{center}
\caption{The cumulative distribution of $P_{LS}$ for the samples
  defined by n=5, 10, 15 and 20. Using the full range of period a KS
  test indicates that they do not come from the same parent
  population.}
\label{madcum} 
\end{figure}

Figure \ref{madcum} shows the cumulative distribution of these
samples. Using the full period range, a KS test suggests that these
distributions do not come from the same parent population. If we apply
the same test using an upper limit of 40 min to the period
distributions we find that the MAD5 and MAD10 samples are consistent
with coming from the same parent population. On the other hand, if we
apply an upper limit of 20 min, we find that all the distributions are
consistent with coming from the same parent population. We therefore
use the MAD5 sample as derived through the LS test as a starting point
for identifying variable stars with periods $<$20 min (since UCBs in
this period range are intrinsically rare) and the MAD10 sample for
periods $>$20 min. In \S \ref{novphas} we use the MAD15 sample to
search for interesting sources which have no VPHAS+ colours.

One consequence of fixing the long period limit of the LS periodogram
to the duration of light curve is that long period variables can be
missed. (This can be eliminated if we set the limit to 20 percent
greater than the duration of the light curve and this will be applied
in the analysis of future data). Instead the AoV test allows the
identification of variables which have a period longer than 2 h and
also eclipsing variables using a phase binning approach
(Schwarzenberg-Czerny 1989, Devor 2005). We have identified a set of
variable candidates based on the AoV test and applying the same `MAD'
process as outlined above. However, in the interests of space we do not
outline a full analysis of these AoV samples here, but later we
highlight three blue variables which were identified through this
test.

In the next section we show that an additional manual verification
stage finds a significant percentage of candidate short period
variables identified using the LS test that are likely false positive
variables. However, in Paper III, we show that followup photometry of
stars which have passed both verification stages confirm the period
determined using OW data, thereby verifying that our pipeline can be
used to identify short period variable stars, including a reliable
estimate of the period.

\section{Searching for Ultra Compact Binaries}

AM CVn binaries together with other compact objects (e.g. single white
dwarfs, cataclysmic variables (CVs)) are located in a distinct region
of the $g-r,u-g$ colour-colour plane (see e.g. Carter et al
2013).\footnote{As the VPHAS+ photometry and the OW $g$ mag are on the
  Vega system, but the SDSS photometry is on the AB system, we
  transform SDSS colours of specific stars onto the Vega systems
  using the equations of Blanton \& Roweis (2007).}  Since the OW
survey obtains observations in the $g$ band only, we obtain colour
information from the VPHAS+ survey (Drew et al. 2014) whose footprint
covers the southern Galactic plane in the same manner as the OW
survey.

Given the rarity of AM CVn stars and other UCBs, the MAD5 sample
(which gives the greatest number of stars, but also the greatest
number of false positives) has been used as the primary sample to
search for blue variable objects. Before the filtering or colour
selection stages there are 31388 objects and after the automatic
flagging procedure there are 26259 objects which remain, with the vast
majority being flagged as they are close to a bright star and hence
their light curves may have been affected by diffraction spikes. We
have used the VPHAS+ second data release (DR2) and made an initial
colour selection of $u-g<$ 0.0 (where we have included stars whose
colour meets this condition within the error on its colour index).
After this colour cut we have a provisional set of 39 blue short
period variable candidates.

Given this is a small number of candidates, we were able to perform a
manual verification procedure which consisted of visually inspecting
the light curves, power spectra and their location on the detector. We
find that of the 39 blue candidate variables, 22 (56 percent) had to
be removed from our list. Most of the stars have been removed because
they appear `swept' by diffraction spikes over the course of the 2 hr
series of observations which affects the brightness of the star and
the local background (see Figure \ref{spikes}). There were several
cases where more than one star was located on the same chip with a
similar shape of light curve indicating a systematic trend was still
present in the data. This finding indicates that there are still a
large number of false positives in the sample selected automatically,
which have not been flagged by the pipeline discussed in \S 3. Work is
on-going to improve the efficiency of the pipeline to flag spurious
variables in an automatic manner. We show the details of the blue variable 
stars which have passed the manual verification stage in Table 3,
the location in the $g-r,u-g$ and $r-i,r-H\alpha$ planes in Figure
\ref{col-col}, and their light curves in Figure
\ref{bluelight}. Finding charts for these stars are shown in Figure
\ref{finding}. 

Of the 17 stars which were identified using the LS test and shown in
Table 3, four have $P_{LS}<40$ min, making them prime UCB
candidates. 

\begin{itemize}

\item OW J074106.0--294811.0, which shows a period of 22.6 min in the
  OW light curve, is one of the bluest variable star
  outlined in Table 3. We presented an optical spectrum of this star
  obtained using SALT in Paper I, which showed H$\alpha$ as a very weak
  absorption line, while H$\beta$ and H$\gamma$ showed stronger
  absorption lines. Further observations of this star have allowed us
  to identify this object as a UCB with an orbital period of 44 m. It
  appears to be a very rare UCB in that it is composed of an sdO/sdB
  star plus an unseen companion (Kupfer et al in prep.).

\item We do not currently have an optical spectrum for OW
  J181038.5--251608.6 ($P_{LS}$=28.9 min) although its colours
  ($g-r$=1.11, $u-g$=--0.50) place it at some distance from the unreddened
  main sequence or white dwarf tracks (Figure \ref{col-col}). There is
  no known X-ray source within 20 arcsec of its optical
  position. 

\item Followup spectra of OW J175358.8--310728.9 ($P_{LS}$=35.2 min)
  indicate that this very blue star is a DQ (carbon rich) white dwarf
  (Macfarlane et al. in prep).

\item We show a spectrum of OW J075527.6--314825.2 ($P_{LS}$=38.6 min)
  in Paper III from which we infer a spectral type of F6/F7. Its
  colours ($g-r$=0.64, $u-g$=--0.18) are slightly displaced from the
  main sequence (it has a bluer $u-g$ colour) which may imply another
  stellar component to the system.

\end{itemize}

As indicated in \S 3 we also identified samples of variable stars using the
AoV phase binning test. For the AoV sample which contained the stars
whose light curve were the statistically most variable, three stars
were bluer than the main sequence and our manual verification
processes confirmed the period found from the AoV test. Although their
periods lie in the range 85--120 min, and are therefore not UCBs, they
show that we can identify short period eclipsing binaries and we
outline their details in Table 3 and Figure \ref{col-col} with their
finding charts shown in Figure \ref{finding}.

In the following sections we discuss these variables together with
some of the other blue variable stars identified in Table 3. In
addition we examine the observational characteristics of the {\delSct}
type stars identified in our survey and some of the previously known
variables which are in the OW fields.

\section{Hot single stars}

\begin{figure*}
\begin{center}
\setlength{\unitlength}{1cm}
\begin{picture}(12,6)
\put(-4.,-0.5){\includegraphics{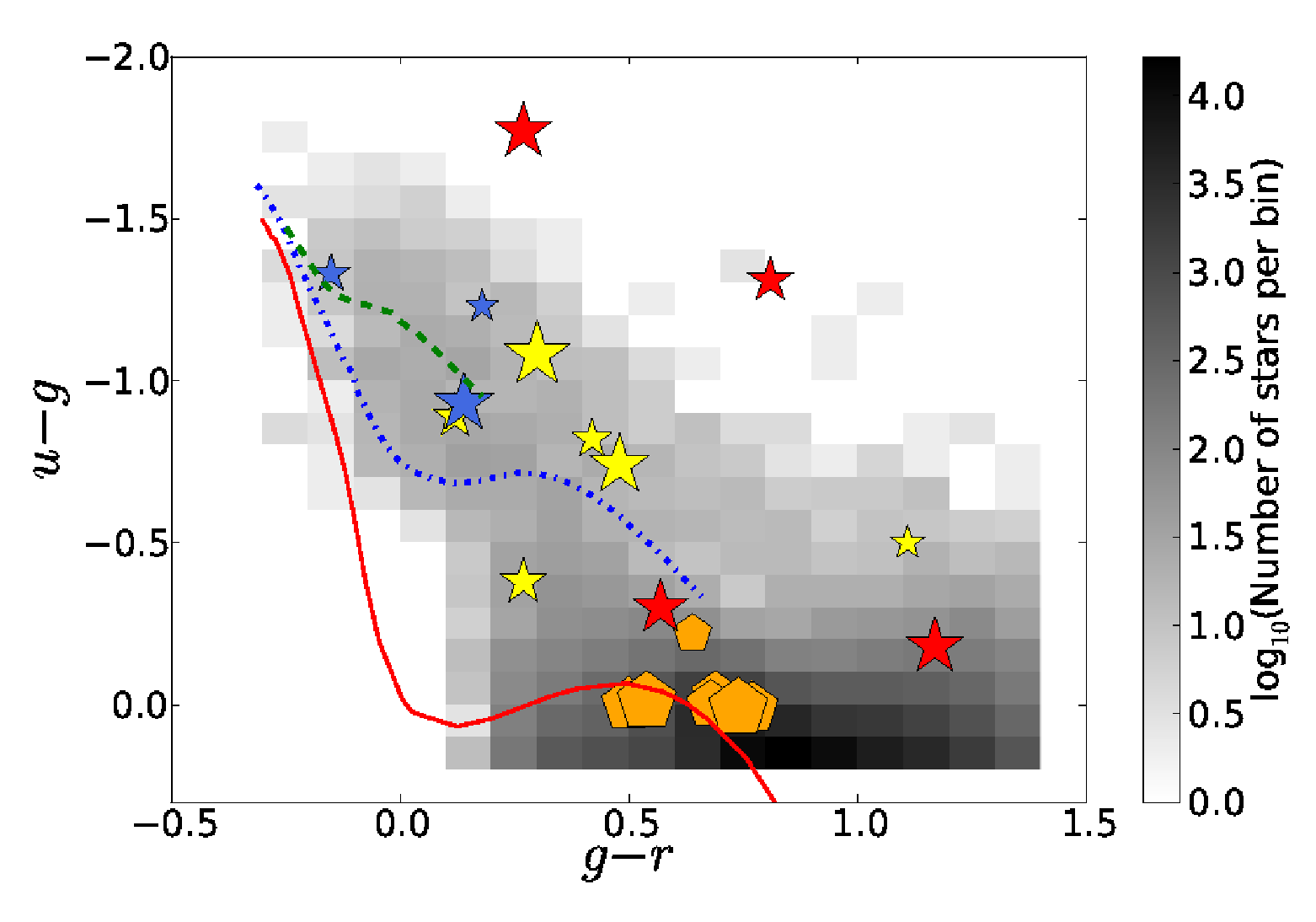}}
\put(5.5,-0.5){\includegraphics{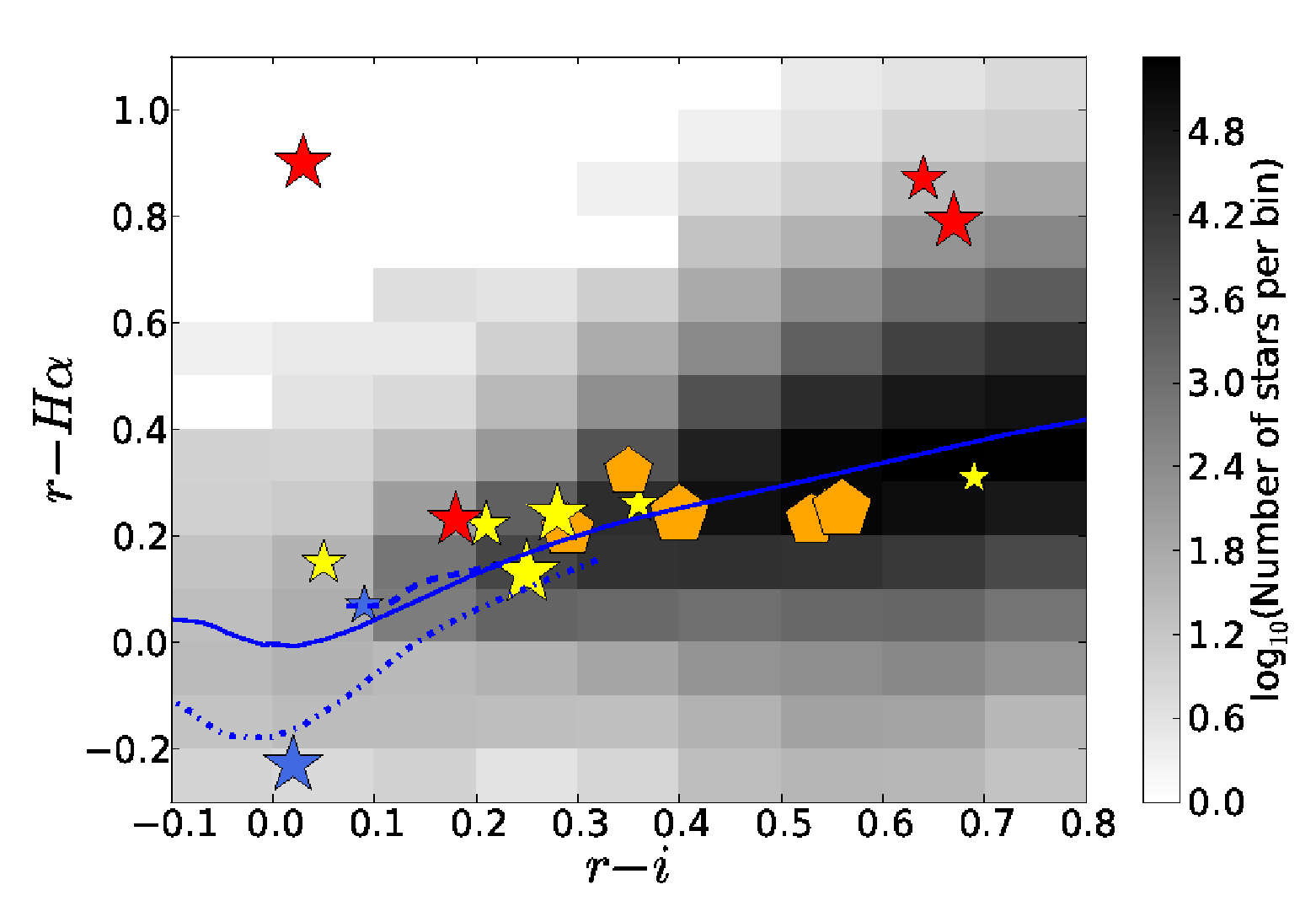}}
\end{picture}
\end{center}
\caption{The $g-r, u-g$ (left) and $r-i, r-H\alpha$ (right) colours of
  the variable stars shown in Table 3. The red line marks the
  unreddened main sequence (taken from Drew et al. 2014), the
  blue line marks the cooling track for DA white dwarfs with log $g$=8 and
  the green line the cooling track for DB white dwarfs with log $g$=8
  (both taken from Raddi et al. 2016). The size of the symbol
  increases as a function of period of the variable.  Red refers to
  CVs, accreting objects or eclipsing binaries; blue represents UCB,
  DQ or DA-like white dwarf; {\delSct} type stars are shown as orange
  polygons and currently unknown objects are yellow stars. The
  pixelated grayscale map shows the number of stars (whether variable
  or not) in the OW survey with VPHAS+ colours.}
\label{col-col} 
\end{figure*}

\begin{figure*}
\begin{center}
\setlength{\unitlength}{1cm}
\begin{picture}(12,12)
\put(2,2){\includegraphics{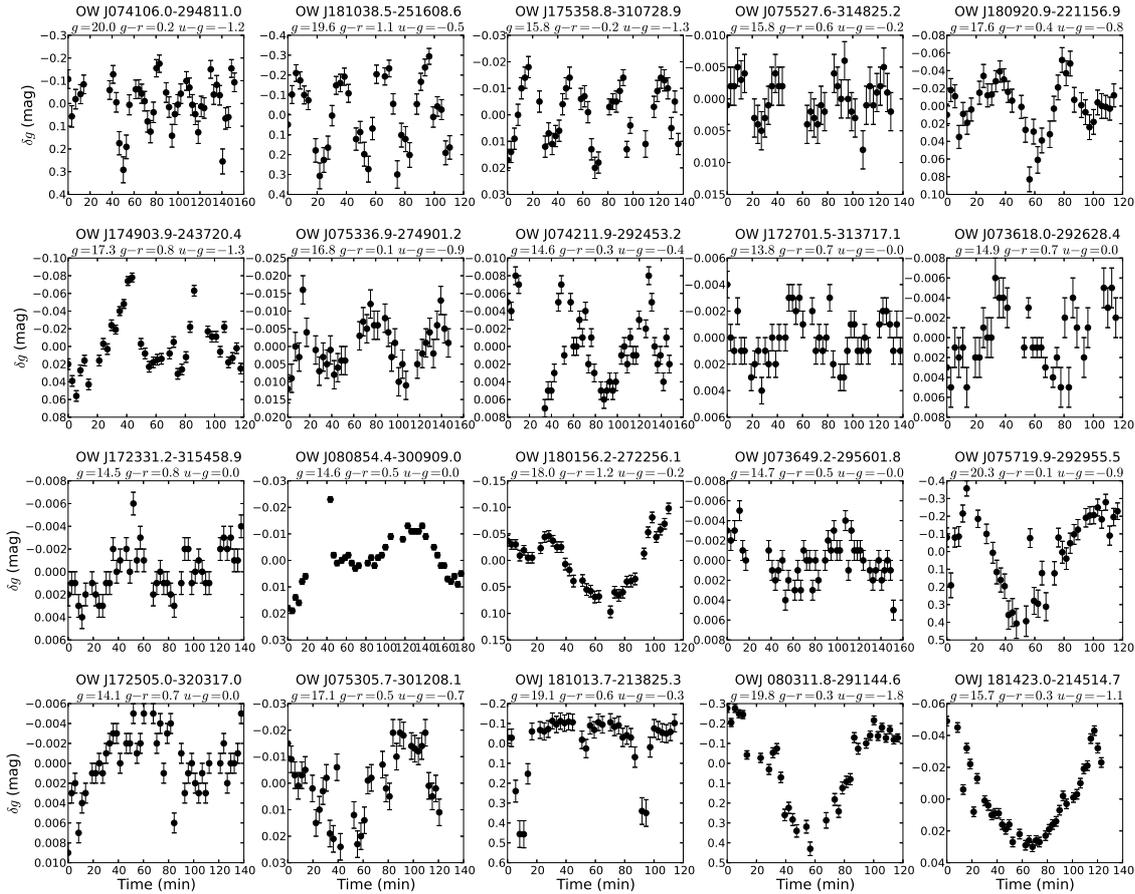}}
\end{picture}
\end{center}
\caption{Light curves of the 20 blue variable stars shown in Table 3
  and whose colours are shown in Figure 7. We show their name, OW$g$
  mag, $g-r$ and $u-g$ colour indices above each light curve.}
\label{bluelight} 
\end{figure*}

OW J075719.9--292955.5 ($P_{LS}$=98.7 min), has very blue colours
($g-r$=0.14, $u-g$=--0.93) but also shows an $r-H\alpha$ index which
implies a deep H${\alpha}$ absorption line (Figure \ref{col-col}). Its
period suggests that it is unlikely to be a pulsating DA white dwarf
but could be a sdO/sdB or GW Vir pulsator. It could alternatively be
the signature of the rotation period of a white dwarf (we show
followup spectra of this star in Paper III).

OW J075305.7--301208.1 ($P_{LS}$=102.2 min), has very blue colours
($g-r$=0.14, $u-g$=-0.93) which are close to the unreddend DA white
dwarf track shown in Figure \ref{col-col}. We show in Paper III its
optical spectrum which is similar to that of an O or a B-type
star. Further spectra with higher signal-to-noise are planned to
determine its nature.

\begin{table*}
\begin{center}
\label{variableblue}
\caption{The 20 blue stars which were identified as variable in the
  MAD5 sample and subsequently passed the manual verification
  stage. We give the sky co-ordinates, the period corresponding to the
  most prominent peak in the LS power spectrum and its corresponding
  False Alarm Probability (FAP), the OW $g$ mag and amplitude of the
  variation, the colours derived from VPHAS+ data and the difference
  between the $g$ mag derived in the OW $g$ data and the VPHAS+ data
  (OWg-Vg). The lower three stars were identified by means of the
  AoV test and give the period and FAP of the most prominent
  period. Notes: (1) PaperI; (2) Kupfer et al. (in prep); (3)
  Macfarlane et al. (in prep); (4) Paper III.}
\resizebox{\textwidth}{!}{
\begin{tabular}{lllrrrrrrrrrr}  
\hline
Name           & RA      & Dec     & $P_{LS}$ & LS & OW$g$ & Amp   & $u-g$ & $g-r$ & $r-i$ & $r-H\alpha$ & OWg- & Type\\
               & (J2000) & (J2000) & (min)   &  FAP   & (mag) & (mag) &       &       &       &             &  VPHASg       &      \\
\hline
 OW J074106.0-294811.0 &  07:41:06.0 &-29:48:11.0 & 22.6 &-3.94   &   20.03 &0.224    &   -1.23  &  0.18 & 0.05  &       & -0.03 &UCB (1,2)\\
 OW J181038.5-251608.6 &  18:10:38.5 &-25:16:08.6 & 28.9 &-4.88   &   19.62 &0.423    &   -0.50  &  1.11 & 0.69  &  0.31 & -0.27 &\\
 OW J175358.8-310728.9 &  17:53:58.8 &-31:07:28.9 & 35.2 &-4.72   &   15.81 &0.026    &   -1.33  & -0.15 & 0.09  &  0.07 &  0.08 &DQ White dwarf(3)\\
 OW J075527.6-314825.2 &  07:55:27.6 &-31:48:25.2 & 38.6 &-3.17   &   15.84 &0.007    &   -0.18  &  0.64 & 0.40  &       & -0.10 &$\delta$ Sct like (4)\\
 OW J180920.9-221156.9 &  18:09:20.9 &-22:11:56.9 & 44.7 &-2.57   &   17.57 &0.056    &   -0.82  &  0.42 & 0.36  &  0.26 &  0.68 &\\
 OW J174903.9-243720.4 &  17:49:03.9 &-24:37:20.4 & 55.6 &-3.28   &   17.31 &0.071    &   -1.31  &  0.81 & 0.64  &  0.87 & -3.44 &CV\\
 OW J075336.9-274901.2 &  07:53:36.9 &-27:49:01.2 & 59.4 &-2.60   &   16.76 &0.014    &   -0.89  &  0.12 & 0.05  &  0.15 &  0.04 &\\
 OW J074211.9-292453.2 &  07:42:11.9 &-29:24:53.2 & 61.1 &-3.14   &   14.62 &0.008    &   -0.38  &  0.27 & 0.21  &  0.22 &  0.16 &\\
 OW J172701.5-313717.1 &  17:27:01.5 &-31:37:17.1 & 63.2 &-2.57   &   13.83 &0.003    &   -0.03  &  0.69 &       &  0.22 &  0.03 &$\delta$ Sct like\\
 OW J073618.0-292628.4 &  07:36:18.0 &-29:26:28.4 & 65.9 &-2.58   &   14.85 &0.006    &    0.00  &  0.68 & 0.35  &  0.32 & -0.06 &$\delta$ Sct like\\
 OW J172331.2-315458.9 &  17:23:31.2 &-31:54:58.9 & 75.3 &-2.75   &   14.54 &0.004    &    0.01  &  0.77 & 0.53  &  0.23 & -0.09 &$\delta$ Sct like\\
 OW J080854.4-300909.0 &  08:08:54.4 &-30:09:09.0 & 84.4 &-2.66   &   14.63 &0.018    &    0.00  &  0.50 & 0.29  &  0.21 & -0.10 &$\delta$ Sct like\\
 OW J180156.2-272256.1 &  18:01:56.2 &-27:22:56.1 & 92.4 &-4.11   &   18.05 &0.128    &   -0.18  &  1.17 & 0.67  &  0.79 & -0.07 &CV, V5627 Sgr\\
 OW J073649.2-295601.8 &  07:36:49.2 &-29:56:01.8 & 95.8 &-3.66   &   14.66 &0.004    &   -0.01  &  0.54 & 0.40  &  0.24 &  0.00 &$\delta$ Sct like\\
 OW J075719.9-292955.5 &  07:57:19.9 &-29:29:55.5 & 98.7 &-3.46   &   20.27 &0.440    &   -0.93  &  0.14 & 0.02  & -0.23 & -0.11 & DA WD/SdB? (4)\\
 OW J172505.0-320317.0 &  17:25:05.0 &-32:03:17.0 & 99.9 &-2.88   &   14.06 &0.006    &    0.01  &  0.74 & 0.56  &  0.25 & -0.07 &$\delta$ Sct like\\
 OW J075305.7-301208.1 &  07:53:05.7 &-30:12:08.1 &102.2 &-3.28   &   17.06 &0.030    &   -0.74  &  0.48 & 0.28  &  0.24 &  0.00 &O type spectrum (4)\\
\hline
Name           & RA      & Dec     & $P_{AoV}$ & AoV & OW$g$ & Amp   & $u-g$ & $g-r$ & $r-i$ & $r-H\alpha$ & OWg- & Type\\
               & (J2000) & (J2000) & (min)   &  FAP    & (mag) & (mag) &       &       &       &             &  Vg      &      \\
\hline
OWJ 181013.7-213825.3 & 18:10:13.7 & -21:38:25.3 & 85.4 & 10.59 & 19.10 & 0.175  &  -0.30 & 0.57 & 0.18 &  0.23 &  1.09 &Possible Eclipsing Binary\\
OWJ 080311.8-291144.6 & 08:03:11.8 & -29:11:44.6 & 90.2 & 15.90 & 19.75 & 0.500  &  -1.77 & 0.27 & 0.03 &  0.90 & -0.02 &Eclipsing CV (3) \\
OWJ 181423.0-214514.7 & 18:14:23.0 & -21:45:14.7 & 118.9 & 17.06& 15.74 & 0.058  &  -1.08 & 0.30 & 0.25 &  0.13 & -0.01 & \\
\hline
\end{tabular}}
\end{center}
\end{table*}

\section{Accreting Binaries}
\label{accrete}

Three of the stars shown in Table 3 and in Figure \ref{col-col} have
colours indicating H$\alpha$ in emission. Two of these have a period
greater than 90 min suggesting that they are typical hydrogen
accreting binaries. A third (OW J174903.9--243720.4) has a period of
55.6 min which, if this could be verified as the orbital period, would
place it well below the CV orbital period minimum, potentially in the
group of helium rich CVs (see Breedt et al. 2014 and references
therein). It also shows a significant difference in brightness between
the epoch of the OW observations and the VPHAS+ epoch ($\Delta g$=3.4
mag) indicating it is an outbursting system. 

OW J180156.3--272256.2 (V5627 Sgr) has strong H$\alpha$ emission but
only moderately blue colours ($g-r$=1.17, $u-g$=--0.18). Indeed, it
was identified as a suspected nova using data taken by the MACHO
project (which aimed to search for dark matter in the form of massive
compact hallo objects, i.e. MACHOs; see e.g. Alcock et al 2000) since
its brightness declined from $V\sim$ 15.1 to 16.9 mag over $\sim$1600
days (see Mr\'{o}z et al. 2015 and references therein). Woudt, Warner
\& Spark (2005) obtained high speed photometry when the star had a
mean $V\sim$ 16.92. Although it showed considerably variability, they
found a stable period of 2.8 h which they took to be the orbital
period. The OW light curve shown in Figure \ref{bluelight} shows a
trend but no evidence of a repeated feature (the duration of the light
curve was less than 2.8 hr). The mean magnitude was OW$g$=18.1
suggesting the brightness of the star has continued to decline from
its outburst.

\begin{figure}
\begin{center}
\setlength{\unitlength}{1cm}
\begin{picture}(6,9.5)
\put(7,4.4){\includegraphics{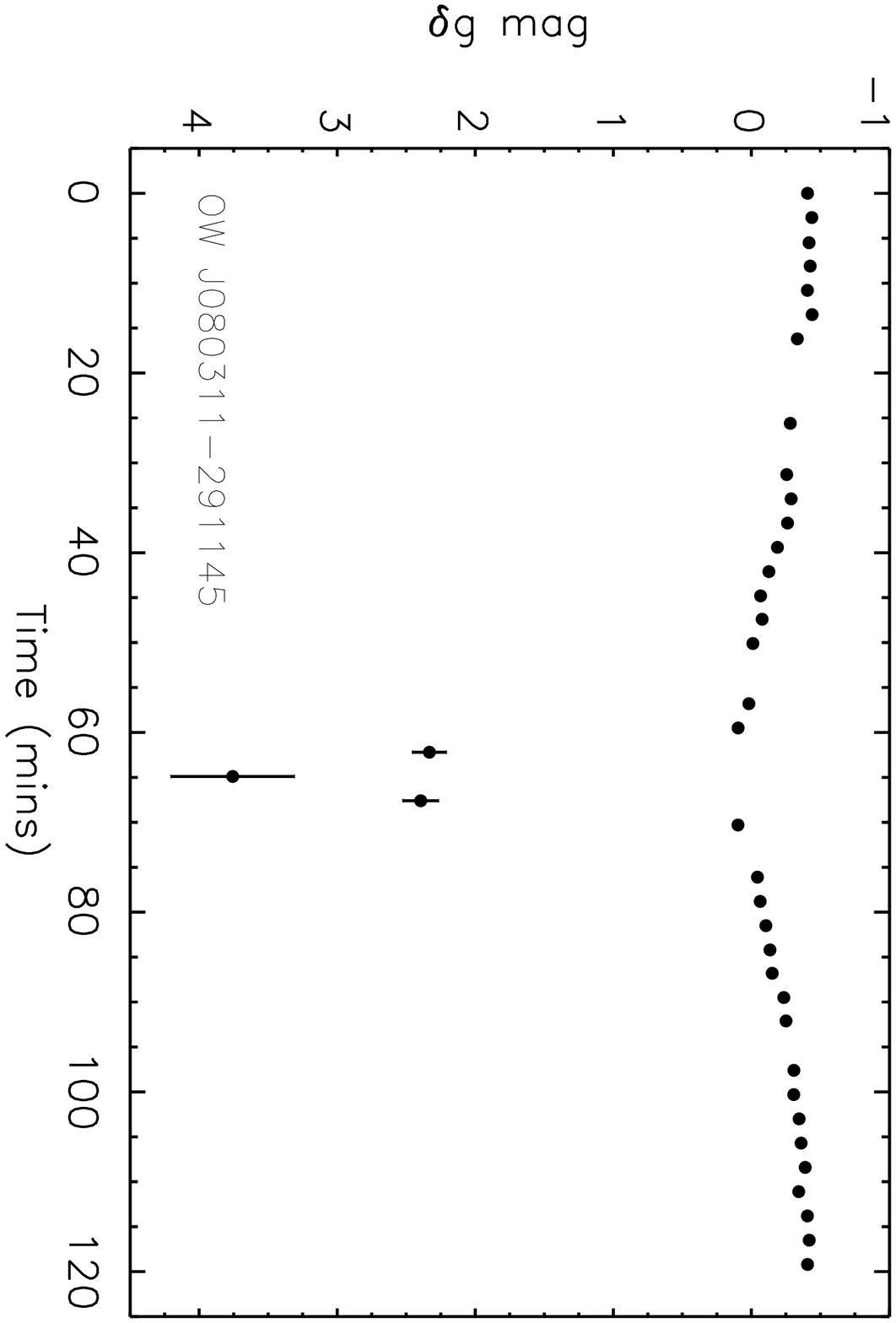}}
\put(7,-0.5){\includegraphics{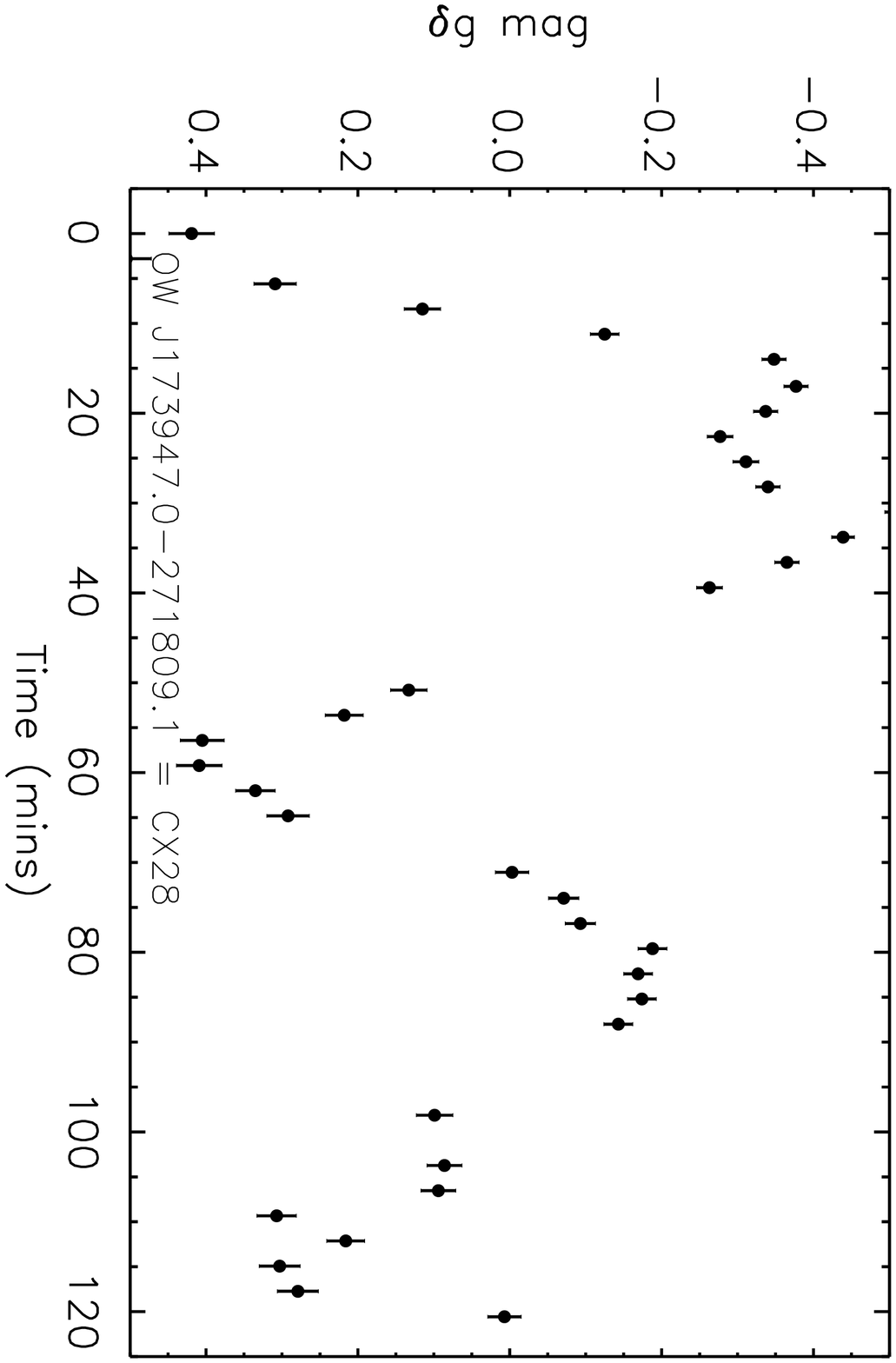}}
\end{picture}
\end{center}
\caption{CVs identified in the OW survey. Top panel: The eclipsing CV
  OW J080311-291145 which was discovered from its OW light curve and
  followup spectroscopy (Paper III); Bottom panel: OW
  J173947.0-271809.1 which was identified as a possible Intermediate
  Polar in the Chandra Bulge X-ray survey (Britt et al. 2013).}
\label{CVs-light} 
\end{figure}

The bluest of these three accreting objects (OW J080311--291145) has a
high amplitude of variability (0.5 mag). The light curve shown in
Figure \ref{bluelight} is derived using the difference imaging
technique that we use in our pipeline. To investigate this source in
more detail we obtained differential aperture photometry using the OW
data. The light curve of this source is shown in the top panel of
Figure \ref{CVs-light} and a clear short duration ($<$10.7 min)
eclipse is seen. (The cadence of our observations preclude a more
precise estimate of the eclipse duration). The reason that the
photometric points during eclipse were not recorded in the
differential imaging light curve is due to the fact that we place a
3$\sigma$ lower limit for detecting stars in an image. (Since this may
have implications for identifying other eclipsing systems, we are
re-assessing this part of the pipeline). Paper III shows its optical
spectrum which confirms this as a high inclination CV.

The OW data implies that the orbital period of OW J0803 is greater
than $\sim$60 min. For comparison the magnetic eclipsing CV UZ For has
an orbital period of 126.6 min and an eclipse duration of $\sim$0.063
cycles (e.g. Bailey \& Cropper 1991). Taking an upper limit of 10.7
min as the duration of the eclipse in OW J080311-291145, this suggests
by comparison with UZ For, an orbital period of $\sim$170 min. If the
eclipse duration was as short as 9 min, this would imply an orbital
period of $\sim$140 min. Although there is a great deal of uncertainty
on the orbital period of this new CV, it is potentially interesting in
that it may reside in the 2--3 h CV orbital period gap (see Zorotovic
et al. 2016 for a recent discussion of the period gap). Photometric
observations are strongly encouraged to determine the orbital period
of this eclipsing CV.

\section{A possible non-interacting, short period, eclipsing system}

In the previous section we showed that OW J080311--291145 has
  been identified as a CV through its VPHAS+ colours and the eclipse
  in its OW light curve. We identified OW J181013.7--213828.3 as a
  variable through the AoV test and find that it shows two
  eclipse-like features with one possible secondary eclipse (Figure
  \ref{bluelight}). If this result can be confirmed, then it would be
  a binary with an orbital period of $\sim$85 min. The VPHAS+ colours
  do not indicate H$\alpha$ in emission and this is confirmed from our
  spectra shown in Paper III.  Although the mean magnitude of OW
  J181013.7 during the OW observations differs by $\sim$1.1 mag
  compared to the VPHAS+ $g$ band magnitude, this is greater than the
  depth of the eclipse like feature seen in the OW photometry. Further
  observations are required to confirm the eclipse-like features and
  search for changes in its mean long term brightness which would be
  an indicator of sporadic mass transfer.

\section{{\delSct}-type Pulsators}
\label{deltascuti}

The seven stars shown in Table 3 and Figure \ref{col-col} have colours
consistent with that of stars whose intrinsic colours are close to
that of main sequence stars of A/F spectral type, or stars with low
reddening, or with intrinsically blue stars which have been heavily
reddended. They all show very low amplitude modulation ($<$0.02
mag). We currently have optical spectra for two of these stars and
they indicate a spectral type F6-F8 (Paper III). The most likely
scenario is that these are {\delSct} type stellar pulsators.

Indeed, surveys with a similar strategy to OW have been successful in
identifying short period stellar pulsators (e.g. Ramsay et al. 2006,
Ramsay et al. 2014). Many of these are $\delta$ Sct type stars with
late-A to mid-F spectral type and dominant periods in the range
of $\sim$26 min to $\sim$6 h (see e.g. Breger 2000 and Chang et
al. 2013). Other types of pulsators with similar spectral type include
$\gamma$ Dor, roAp and SX Phe stars and only medium resolution spectra
with high signal-to-noise can distinguish the specific class of
variable star. 

\begin{figure}
\begin{center}
\setlength{\unitlength}{1cm}
\begin{picture}(12,6)
\put(10.5,-1){\includegraphics{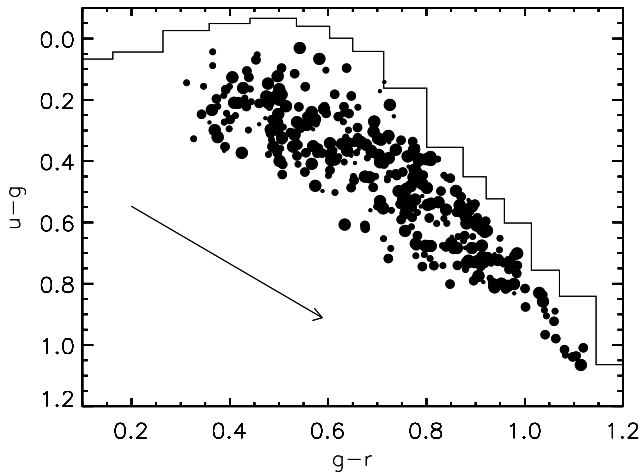}}
\end{picture}
\end{center}
\caption{The $g-r,u-g$ colours of the $\delta$ Sct type variables
  where the size of the symbol increases as a function of period. The
  arrow represents the direction and magnitude of the reddening vector
  for $A_{V}$=1.0 and the histogram represents the unreddened main
  sequence (from Drew et al. 2014).}
\label{col-col-delta-sct} 
\end{figure}

To identify a larger sample of candidate $\delta$ Sct type pulsators
from the OW survey we took the MAD10 sample as an initial starting
point and searched for stars with a period less than 1 hr,
approximately half the duration of the OW light curve.  Given that the
fields are close to the Galactic plane, we expect many stars will have
colours which are significantly reddended (in contrast to the seven
stars which are likely {\delSct} type stars shown in Table 3). In
selecting a sample of {\delSct} candidates, we assume they have
intrinsic colours in the range A0 and F5. We therefore select all
stars which have a $u-g$ colour redder than the main sequence and
$u-g, g-r$ colours which are consistent with intrinsic colours of
A0--F5 stars. This is best seen in Figure \ref{col-col-delta-sct}
which shows our sample of $\delta$ Sct type stars in the $g-r, u-g$
plane.

\begin{figure}
\begin{center}
\setlength{\unitlength}{1cm}
\begin{picture}(12,6)
\put(10.5,-1){\includegraphics{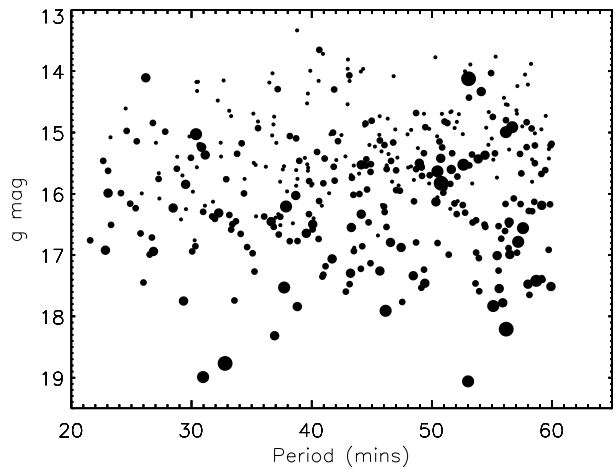}}
\end{picture}
\end{center}
\caption{The $\delta$ Sct type pulsators which have been selected due
  to their colour and variability as a function of period and $g$
  mag. The size of the symbol reflects the amplitude of the
  variability ($<$0.01 mag for the smallest symbols, with $>$0.05 mag
  for the largest symbols).}
\label{amp-per-delta-sct} 
\end{figure}

There are 5263 variables in the MAD10 sample which have a period of
$P_{LS} <$ 60 min and passed our initial flagging stages. Of these 455
have VPHAS+ colours which are consistent with a star with intrinsic
colours of an A0 to F5 star. The significantly smaller number of
candidates with both colour indices within the expected range is
consistent with the relatively low values of the percentages of stars
with colour information (for instance only 23 percent of the 5263
stars currently have a $u-g$ index) due to the incompleteness of the
VPHAS+ dataset which is still being obtained. After a manual
verification phase we removed 78 stars out of 455 stars (18 percent)
because their light curves were affected by very close stars or are in
the diffraction spikes of a nearby bright star. We show the period and
OW$g$ mag of the remaining stars in Figure \ref{amp-per-delta-sct} where
the symbol size reflects the amplitude of variability. Of the 377
stars, 39.8 percent stars have an amplitude less than 0.01 mag, 1.1
percent have an amplitude greater than 0.05 mag, the highest amplitude
being 0.11 mag. The star with the shortest period has a period of 9.3
min. We discuss the characteristics of this sample in \S
\ref{disscussion-delsct}.

\section{Stars with no VPHAS+ colour information}
\label{novphas}

Since the VPHAS+ survey is on-going, many of our variable stars do not
currently have colour information. For instance, for the variables in
the MAD15 sample only 24.0 percent currently have $u-g$ {\sl and}
$g-r$ colours. A set of light curves of short period stars which have
no colour information were selected from the MAD15 sample in order to
detect potentially new interesting stars which are statistically very
likely to be intrinsically variable. The MAD15 sample contains 3168
stars after automatic flagging (c.f. Table 2). Out of these, 2089
stars have no VPHAS+ $g$ band information and of these 227 have
$P_{LS} < $ 20 minutes and 260 have 20 $< P_{LS} < $ 40 minutes.

\begin{figure}
\begin{center}
\setlength{\unitlength}{1cm}
\begin{picture}(12,5.7)
\put(9.6,-0.8){\includegraphics{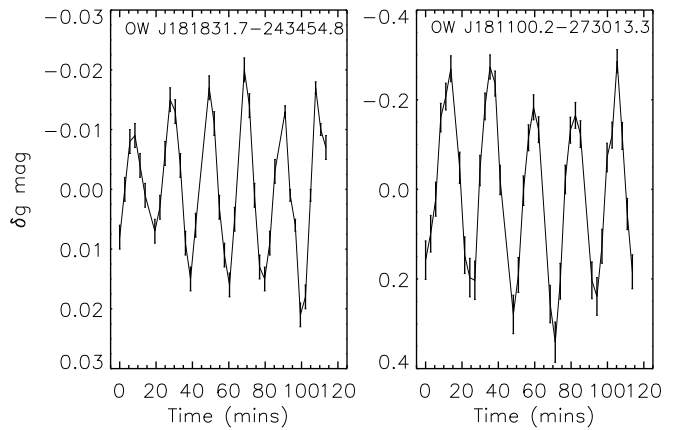}}
\end{picture}
\end{center}
\caption{Two examples of short period pulsating candidates which
  currently have no colour information, left panel OW
  J181831.7-243454.8 (20.4 min) and right panel OW J181100.2-273013.3
  (23.1 min).}
\label{pulse-no-col} 
\end{figure}

We show two examples of short period variables in Figure
\ref{pulse-no-col}. OW J181831.7--243454.8 (OW$g$ = 14.90) has a period
of 20.4 min and an amplitude of 0.029 mag, whilst OW
J181100.2--273013.3 (OW$g$ = 18.40) has a period of 23.1 min and an
amplitude of 0.46 mag. The brighter of these two stars has an
amplitude consistent with $\delta$ Sct type pulsators at the brighter
end of the distribution shown in Figure \ref{amp-per-delta-sct}. The
stars have a very similar period but the fainter of the two has a
much higher amplitude of modulation. Indeed, it is much greater than
the $\delta$ Sct type pulsators shown in Figure
\ref{col-col-delta-sct} making it an interesting object for further
investigation.

Over the course of the VPHAS+ project (Drew et al. 2014) further sky
coverage will be obtained and OW fields which do not currently have
colour information will be matched with these additional new VPHAS+
fields. In the mean time, by comparing the period, $g$ mag brightness
and amplitude of variable stars with no colour information to those which
do have colour information, we will be able to optimally target
followup photometry and spectroscopy for stars which have unusual
characteristics.

\section{Previously known objects in OW fields}
\label{simbad}

To determine which OW variable objects have information in the
literature we took our MAD5 sample after automatic filtering and
searched for objects in the {\tt SIMBAD} database using a 2 arcsec
radius (although the mean uncertainty in our positions is typically
0.1 arcsec, applying a larger search radius allows for a small amount
of proper motion and uncertainty in the position recorded in the {\tt
  SIMBAD} database). We found a total of 179 stars out of the 26260 in
the {\tt SIMBAD} data base. The greatest proportion of stars are RR
Lyr stars (76/179). Given that the photometric period of these stars
is much longer (typically 0.2--1d) then the duration of the OW light
curves, it is likely that we are catching only a small fraction of the
pulsation period. We also find two CVs, two symbiotic binaries and two
planetary nebulae which we now discuss in more detail.

\subsection{Cataclysmic Variables}
\label{CVs}

There are two known CVs in our MAD5 sample: OW J173947.0--271809.1 and
OW J180156.3--272256.2 (V5627 Sgr).  We discussed V5627 Sgr in \S
\ref{accrete}, while OW J173947.0--271809.1 was identified as an X-ray
source (CX28) in the Chandra Bulge X-ray survey (Jonker et
al. 2011). Its X-ray properties, coupled with its optical spectrum
(Britt et al. 2013), hints at an Intermediate Polar nature (CVs whose
accretion disk is truncated and have a white dwarf magnetic field
strength B$\sim$10 $^{5-7}$ G). The optical photometry shown in Britt
et al. indicates that CX28 is optically variable, with a mean
brightness of $g\sim$ 16.7, a high ($\sim$0.7 mag) amplitude of
variations and some (but not conclusive) evidence of a 2.76 hr
period. We show the OW photometry of CX28 in Figure \ref{CVs-light}
where there is some suggestion of a period of $\sim$1 hr. The star is
also significantly fainter at the time of the OW observations
(OW$g\sim$19.0).

\subsection{Central Stars of Planetary Nebulae}
\label{pps}

Searches for variability in the flux of the central stars of Planetary
Nebulae (CSPN) have been on-going for decades (e.g. Bond 1979,
Miszalski et al. 2009). It is of interest because the observed
distribution of orbital periods can be compared to theoretical models
of the formation of PN which predict their period distribution, in
particular bipolar/asymmetric PN which are predicted to be formed
(perhaps largely) due to binary interactions (e.g. Bond 2000).

Our MAD5 sample includes two stars classified as (possible) Planetary
Nebulae. OW J1722353--3214036 (= Th 3--8) shows a modulation on a
period of 22.6 min. However, this period could not be verified using
differential aperture photometry, perhaps because there is a star in
the wings of its PSF. PN Th 2--8 (OW$g$=15.7) is classed in {\tt SIMBAD}
as a planetary nebula, but Acker et al. (1987) finds no emission lines
in its optical spectrum casting considerable doubt on this claim (a
finding confirmed by our spectrum which is shown in Paper III).

The other PN-like star is OW J1747339--2147231 (= H 2--22) which
shows a partial sinusoidal modulation (OW$g$=19.2, amplitude $\sim$0.2
mag) over its 2 h light curve. The CSPN with the shortest period
appears to be 3.4 h (Miszalski et al. 2009). If the optical
modulation of PN H 2--22 can be confirmed this may place it at the
short period end of the CSPN distribution. Further optical photometry
of this object is clearly desirable.

\subsection{Symbiotic Binaries}

Symbiotic stars are interacting binary systems containing a red giant
star and a hotter component, which is typically a white dwarf (see
Mikolajewska 2007 for a review). Recently evidence has accumulated to
suggest that symbiotic stars could be progenitors of a fraction of
supernovae Ia explosions (e.g. Dilday et al. 2012).  Two of our
variable candidates are classed as symbiotic binaries in the {\tt
  SIMBAD} database. The symbiotic binary SS73 122 (OW
J1804412--2709124) shows a periodic modulation on a period of 22.8 min
and an amplitude of 0.006 mag (OW$g$=15.1). We obtained differential
aperture photometry of this star but were not able to verify the
22.8 min modulation, perhaps because there is a $g$=18.6 mag star
$\sim$6 arcsec distant. Further photometry of this star is required.

\begin{figure}
\begin{center}
\setlength{\unitlength}{1cm}
\begin{picture}(6,5.5)
\put(8,-0.8){\includegraphics{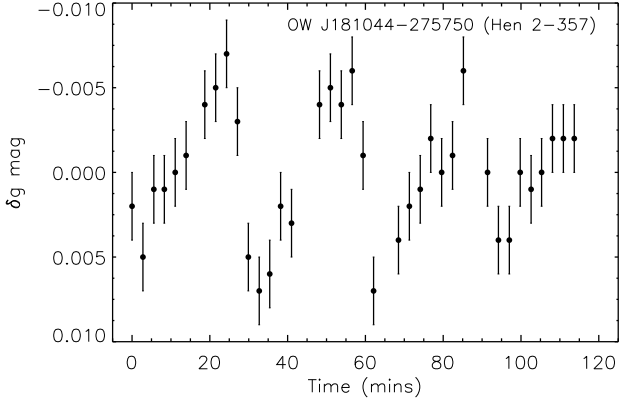}}
\end{picture}
\end{center}
\caption{The OW light curve of the Symbiotic Binary Hen 2--357 which
  shows evidence of a periodic modulation on a period of 31.4
  min. This may imply that it has a significant magnetic field which
  would make it only the second known magnetic Symbiotic Binary.}
\label{Sym-light} 
\end{figure}

The second star, Hen 2--357 (OW J1810439--2757500), shows a modulation
on a period of 31.4 min and an amplitude of 0.008 mag (OW$g$=14.7). This
period is confirmed using differential aperture photometry and we show
its light curve in Figure \ref{Sym-light}. This would make it only the
second known symbiotic binary to show a periodic modulation on a
period less than one hour. The most obvious origin of the modulation
is a hot accretion region on the white dwarf coming into and out of
view as the white dwarf rotates on a period of 31 min (a time-scale
similar to the rotation period of the white dwarfs in CVs). An
alternative scenario is that the 31 min is a quasi-periodic
oscillation due to flickering originating from the accretion disk. A
much longer series of photometry is required to determine how stable
the period is. The optical spectra of both these stars is presented
in Paper III and are consistent with spectra of known symbiotic
binaries, although there is some evidence for extended emission.

The first symbiotic star to be identified as magnetic was Z And
(Sokoloski \& Bildsten 1999) which has an optical light curve which
shows a modulation on a period of 28 min.  Since then only one other
symbiotic system has shown a periodic modulation (BF Cyg at 1.8 h,
Formiggini \& Leibowitz 2009). It is not clear if the lack of
symbiotic binaries showing evidence for a magnetic field is due to the
coherent signal of the rotating white dwarf being greatly diluted by
the light from the mass donating giant star, or that it is an
indication of their formation (c.f. Yungelson et al. 1995).

\section{Discussion}

We have identified samples of variable stars, some of which contain
blue compact stars, and others are likely {\delSct} type pulsating
stars. To determine the nature of specific variable stars, we require
followup photometry and spectroscopy of individual objects, as we have
done in Paper III.

\subsection{The distribution of variable stars}

\begin{figure}
\begin{center}
\setlength{\unitlength}{1cm}
\begin{picture}(12,19)
\put(0.5,10.7){\includegraphics{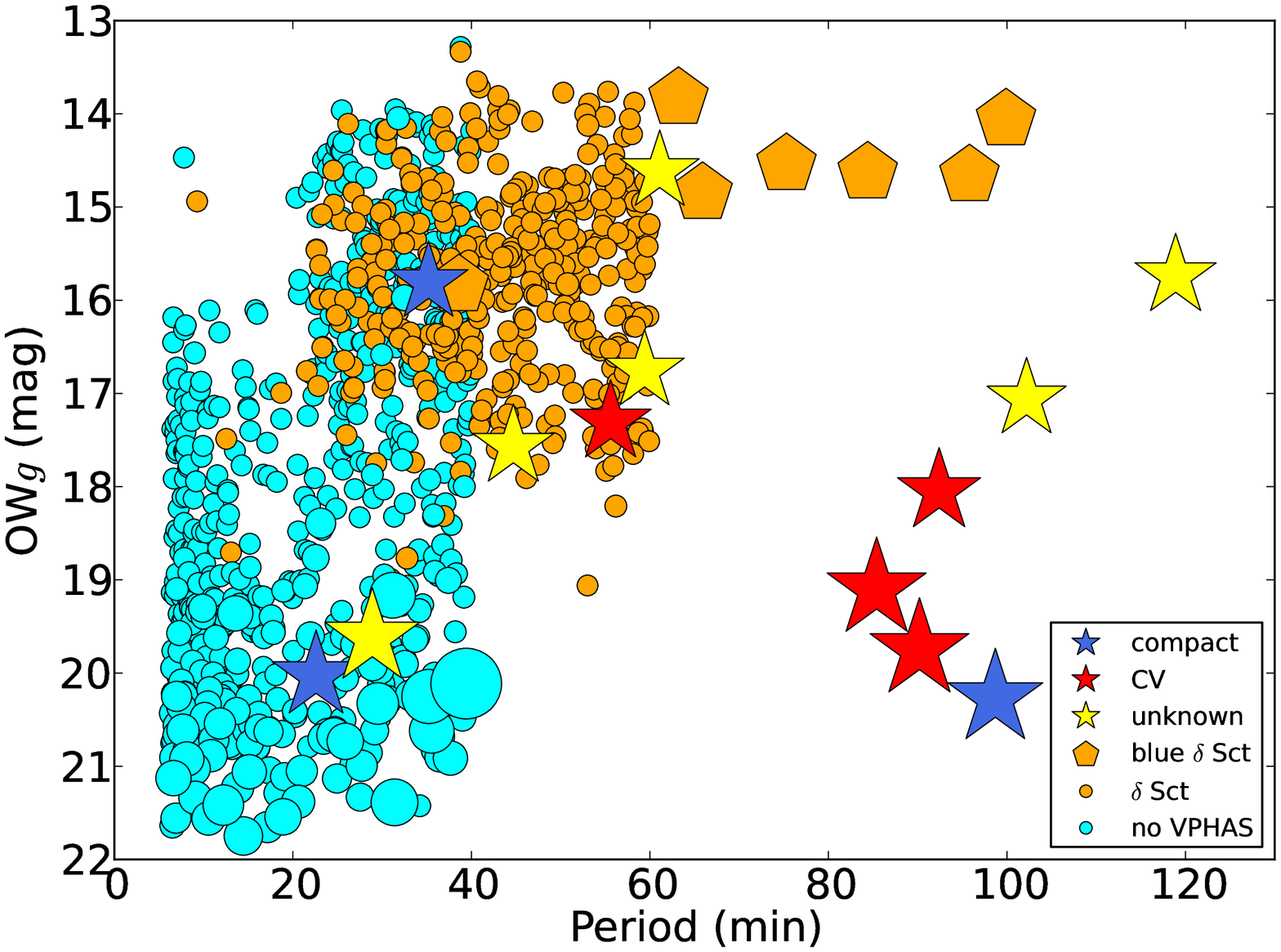}}
\put(0.5,4.4){\includegraphics{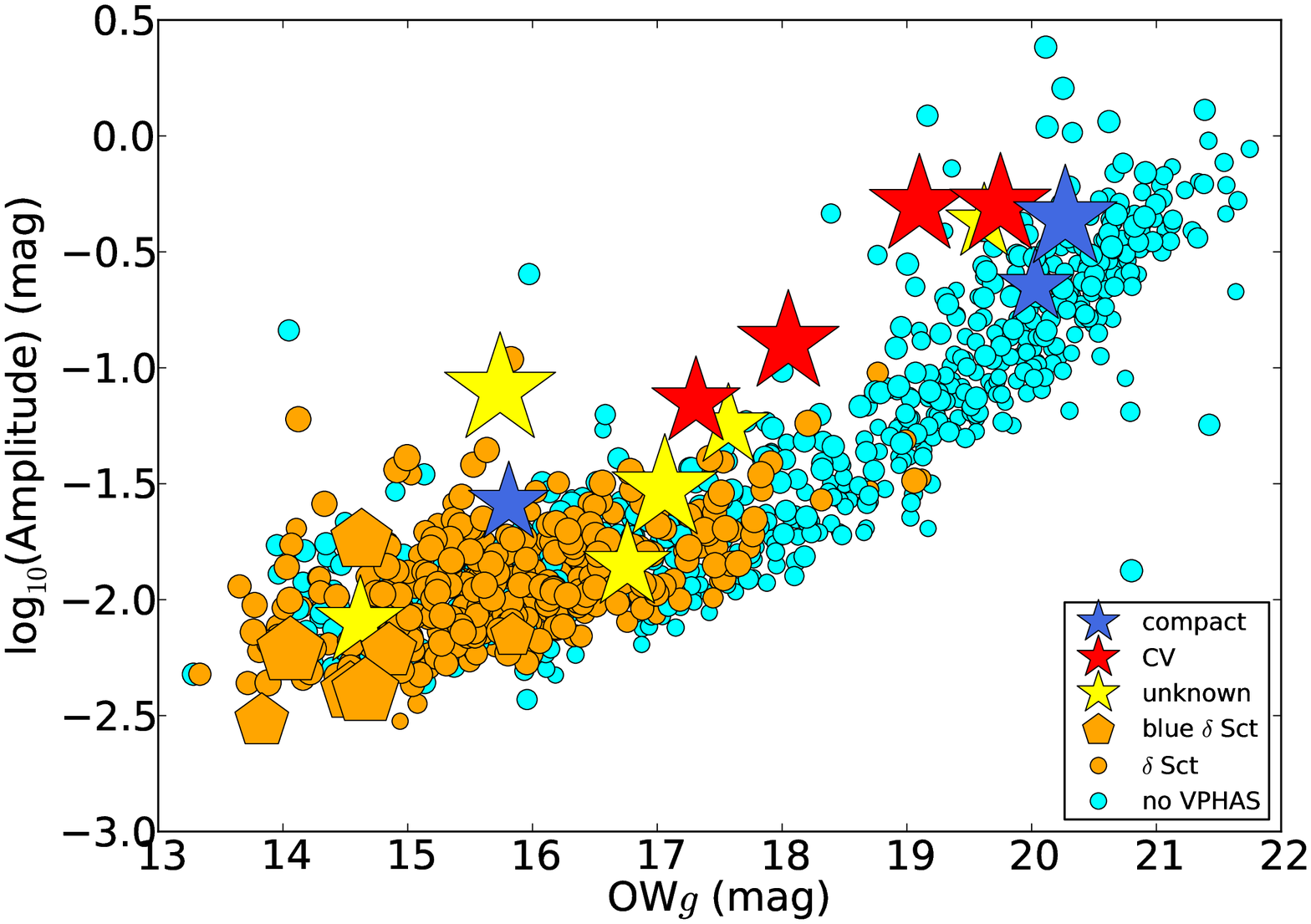}}
\put(0.5,-2.2){\includegraphics{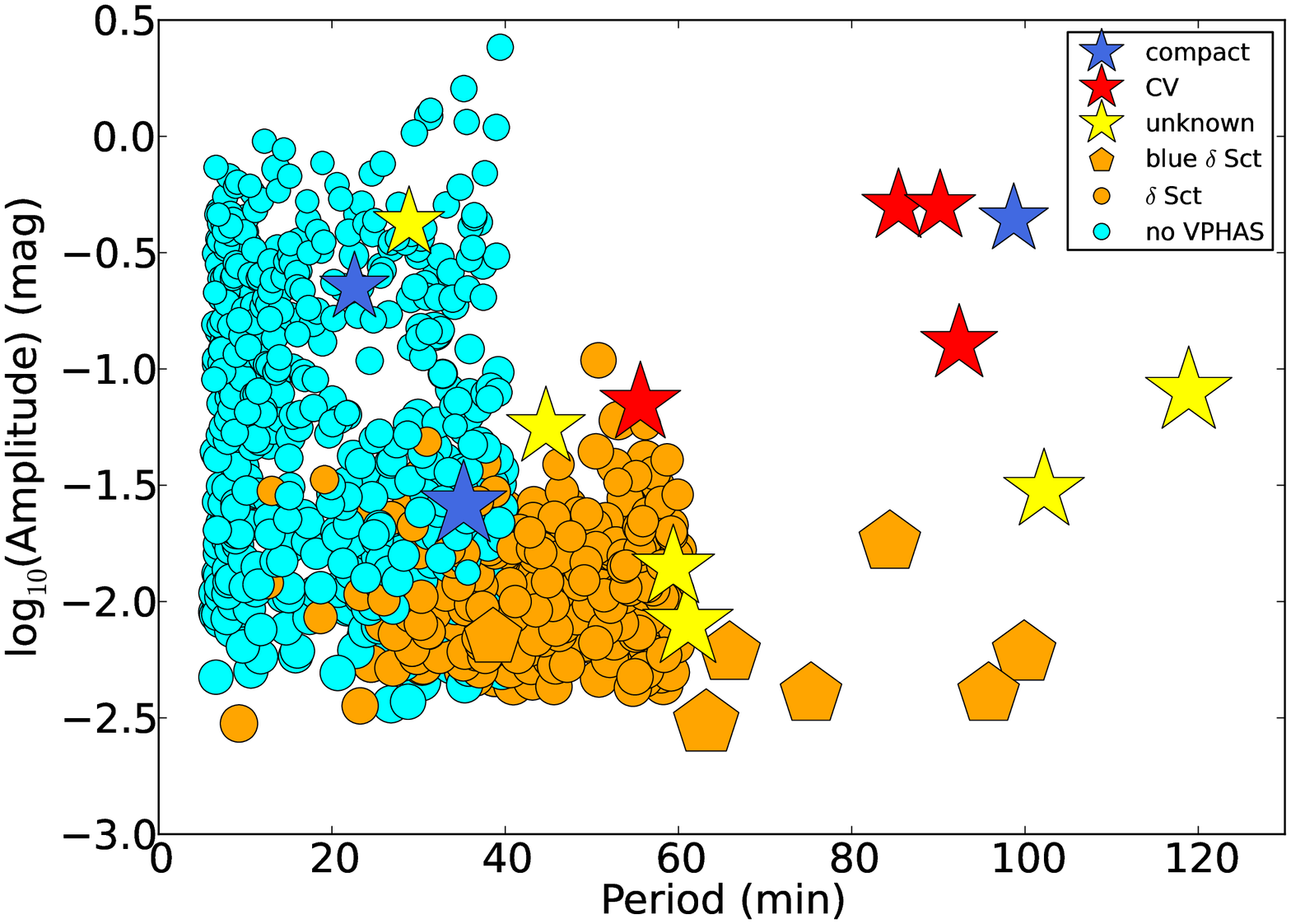}}
\end{picture}
\end{center}
\caption{The distribution of different classes of variable stars in the
  $P_{LS}$, OW$g$ plane (top panel), the OW$g$, Amplitude plane (middle
  panel) and $P_{LS}$, Amplitude plane (bottom panel). Variable stars
  shown in Table 3 and Figure 7 are indicated by the following
  symbols: blue stars refer to systems with a white dwarf; red stars
  indicate CVs, accreting objects or eclipsing binaries; yellow stars
  are variables with currently unknown type, and orange polygons refer
  to {\delSct} stars. Orange circles refer to {\delSct} stars outlined
  in \S \ref{deltascuti} and turquoise circles refer to stars which
  currently have no VPHAS+ information (\S \ref{novphas}). In the
  upper panel the size of the symbol increases as a function of
  amplitude, in the middle it increases as a function of period and
  in the lower panel it increases as the star gets fainter.}
\label{distributions} 
\end{figure}

In Figure \ref{distributions} we show the distribution of different
classes of variable stars in the Period vs OW$g$ magnitude, OW$g$ vs Amplitude, and the
Period vs Amplitude planes. We find that the blue variables shown in
Table 3 which are not {\delSct} types tend to be fainter than
OWg$\sim$16 mag and have $P_{LS}>$ 20 min. The {\delSct} types
typically have $P_{LS}>$ 30 min and are brighter than OWg$\sim$17. On
the other hand, the stars which are either CVs or eclipsing show
amplitudes which are greater than the mean. More compact blue stars
show amplitudes which are more typical of other variables making the
OW$g$ vs Amplitude plane less suitable for identifying them. Many of
the stars which show short periods and do not have VPHAS+ colour
information show a wide range of amplitude. However, the fact that the
short period blue objects also show high amplitudes indicates that
this is a population well worth exploring for new interesting objects.

\subsection{The number of compact stars}

In its first four years, OW has covered 134 square degrees of sky
close to the Galactic plane. Only 45.4 square degrees currently have
$u$ and $g$ band information and only 18.4 percent of the MAD5 sample
of variables have $u-g$ and $g-r$ colour indices. The number of stars
in the $g-r$, $u-g$ colour-colour plane is illustrated in Figure
\ref{col-col} together with the cooling track for DA and DB white
dwarfs with log $g$=8. Around 900 stars in the OW survey are currently
located close to these cooling tracks or along the reddening vectors
implying 20 stars per square degree. This is approximately twice the
number found in the first 211 square degrees of the UVEX survey where
2170 `UV excess' stars were identified (Verbeek et al. 2012).

A detailed search for DA white dwarfs found between 153 and 211
objects in 211 square degrees of the UVEX survey depending on how the
samples were identified (Verbeek et al. 2013). The fact that DA white
dwarfs make up a relatively small percentage of stars in the same part
of the colour-colour plane is not a surprise but does indicate that
spectroscopic followup is essential in determining the nature of any
blue variable star. Our search for variable stars identified around
ten stars which are close to the DA and DB cooling track, giving 0.2
variables per square degree.

Differences between the numbers of compact blue objects derived from
the UVEX and OW surveys could be due to the fact that the initial
release of UVEX data avoided the Galactic Bulge regions, where as
around half of the OW fields have been directed towards the
Bulge. These issues will be explored in future work.

\subsection{The period distribution of $\delta$ Sct stars}
\label{disscussion-delsct}

Chang et al. (2013) provide an overview of the observational
properties of the then known {\delSct} stars -- their catalogue
includes chemically peculiar {\delSct} stars such as $\lambda$ Boo
stars and roAp stars. The surveys which form the basis of this study
have different depth and cadence to ours which will result in built-in
biases. One example of this is the bimodal distribution of the $V$ mag
of their sample which is due to the brighter objects being derived
from the {\sl Hipparcos} catalogue (Perryman et al. 1997, the product
of the ESA's Space Astrometry Mission) and fainter objects from the
MACHO (Alcock et al. 2000) and OGLE (Optical Gravitational Lensing
Experiment, Wo{\'z}niak et al. 2000) surveys.

OW has its own biases -- for instance we have not attempted to include
{\delSct} stars with a pulsation period greater than $\sim$1 hr. In
addition we have optical spectra of only 43 out of the 377 variables
which we have identified as candidate {\delSct} stars. In Paper III we
determine the spectral type of these stars and find they have F
spectral types which are consistent with that of {\delSct} stars.

We highlight two questions here. What is the shortest period and what
is the amplitude distribution of the {\delSct} stars in our sample?
The shortest period {\delSct} stars in the Chang et al. (2013)
catalogue have a period of 25.9 min, although we note that the
pre-main sequence {\delSct} star HD 34282 (Amado et al. 2004) has a
period of 18.1 min. The shortest period {\delSct}-like star we detect
(OW J075633.0-293226.6) has a period of 9.3 min which is shorter than
those noted in Chang et al. (2013). Followup photometry and
spectroscopy of this star is required to better characterise its
properties.

Chang et al. (2013) adopt the same criteria for defining low amplitude
{\delSct} stars as Solano \& Fernley (1997), namely $\delta V$=0.1 mag
with high amplitude {\delSct} stars (HADS) having $\delta V>0.3$
mag. We show the percentage of stars in the Chang et al. (2013)
catalogue which have an amplitude less than 0.01 mag, 0.05 mag and
0.10 mag in Table 4. We do this for the whole sample of Chang et
al. (2013) and also for the (much smaller) sample of stars in the
range 14.5 $<V<$ 20.5 which is more comparable to OW. Whilst the Chang
et al. (2013) sample shows many {\delSct} stars with amplitudes
greater than 0.1 mag, we find {\sl one} star with an amplitude $>$0.1
mag. It is possible that we are identifying very low amplitude
variables of the sort observed using {\sl Kepler} (e.g. Uytterhoeven
et al. 2011). Bowman et al. (2016) find only two {\delSct} stars out
of 983 which were found to be HADS.  (In Paper III we show that
followup observations of our {\delSct} sample indicate that very
similar periods and amplitudes are observed at a second epoch of
observation).

\begin{table}
\begin{center}
\begin{tabular}{llrr}
\hline
Range & Chang & Chang           & OW\\
(mag) & all   & (14.5$<V<$20.5) & \\
\hline
\ltae 0.01  & 9.9\%  & 3.1 \% & 39.8\% \\
\ltae 0.05  & 34.8\% & 21.9\% & 98.9\% \\
\ltae 0.10  & 47.4\% & 37.4\% & 99.7\% \\
\hline
\end{tabular}
\caption{We compare the percentage of {\delSct} stars with an
  amplitude less than three limits in three samples. We take all of
  the sample of Chang et al. (2013), the sample of Chang et al. in the
  14.5 $<V<$ 20.5 range, and the OW sample which has been based on
  colour and period.}
\end{center}
\label{deltaSctamp}
\end{table}

\section{Conclusions}

We have been able to identify a sample of blue stars which show
variability in their light curve on a time-scale as short as $\sim$20
min. Using an automated flagging procedure and subsequent manual
inspection, we find that many of these stars are variable because
diffraction spikes appear to rotate over the detector (and across the
PSF of nearby stars) over the course of the 2 h observation as the
telescope has an alt-az mount.  Although the fraction of contaminants
is high, followup observations reported in Paper III show that our
manual inspection is sufficiently robust that the sample of variable
stars which pass this step are intrinsically variable on the
time-scale identified in our pipeline. The underlying false positive
rate is therefore very low. Our goal is to obtain high cadence light
curves with a duration of 2 h over 400 square degrees close to the
Galactic plane. As we expect UCBs to be rare, we will only be in
  a position to robustly determine the observed space density of short
  period UCBs once the survey approaches the target sky coverage.

\section{Acknowledgements}

The authors gratefully acknowledge funding from the Erasmus Mundus
Programme SAPIENT, the National Research Foundation of South Africa
(NRF), the Nederlandse Organisatie voor Wetenschappelijk Onderzoek
(the Dutch Organization for Science Research), Radboud University and
the University of Cape Town. The ESO observations used in this paper
are based on observations made with ESO Telescopes at the La Silla
Paranal Observatory under programme IDs: 088.D-4010(B), 090.D-0703(A),
090.D-0703(B), 091.D-0716(A), 091.D-0716(B), 092.D-0853(B),
093.D-0937(A), 093.D-0753(A), 094.D-0502(A), 094.D-0502(B), and
177.D-3023 (VPHAS+). This research has made use of the SIMBAD
database, operated at CDS, Strasbourg, France. This research was made
possible through the use of the AAVSO Photometric All-Sky Survey
(APASS), funded by the Robert Martin Ayers Sciences Fund. Armagh
Observatory is core funded by the Northern Ireland Government.  This
research has been facilitated by the NWO-NRF bilateral agreement got
astronomical collaboration between The Netherlands and the Republic of
South Africa. TRM acknowledges the support of STFC through a
Consolidated Grant (ST/L000733).

\appendix

\section{Tables}

\begin{table*}
\begin{center}
\caption{The observation log of field pointings made over ESO Semester 90--94, where 
we show the sky co-ordinates in equatorial and Galactic co-ordinates; the calendar date
of the start of the observations and the range in the seeing in arcsec.}
\label{fieldcentres}
\begin{tabular}{lccrrcc}
\hline
Field & RA      &  DEC    & $l$ & $b$ & Date       & Seeing ($^{''}$) \\
      & (J2000) & (J2000) &     &     & (dd-mm-yy) & Mean, $\sigma$\\
\hline
\multicolumn{7}{l}{Semester P90} \\
\hline
 29a & 07:48:15.9 & --29:39:20.1 & 245.48 & --2.05 & 07--12--2012 & 0.79 0.43 \\
 29b & 07:52:52.1 & --29:39:15.9 & 245.98 & --1.19 & 07--12--2012 & 0.79 0.43 \\
 31a & 07:29:34.1 & --29:41:07.4 & 243.52 & --5.61 & 04--03--2013 & 0.79 0.43 \\
 31b & 07:34:10.3 & --29:41:02.6 & 243.99 & --4.73 & 04--03--2013 & 0.80 0.44 \\
\hline
\multicolumn{7}{l}{Semester P91} \\
\hline 
   6a & 17:02:19.1 & --27:59:57.3 & 355.42 &  8.40 & 01--08--2013         & 1.19 0.55 \\ 
   6b & 17:06:50.8 & --28:00:02.7 & 356.02 &  7.60 & 01--08--2013         & 1.17 0.54 \\ 
  12a & 07:06:21.4 & --30:00:02.6 & 241.51 &--10.25 & 06--04--2013        & 0.85 0.45 \\ 
  12b & 07:10:58.5 & --29:59:57.4 & 241.95 & --9.34 & 06--04--2013        & 0.90 0.47 \\ 
  13a & 17:15:43.4 & --29:59:57.2 & 355.54 &  4.85 & 16--05--2013         & 1.26 0.54 \\ 
  13b & 17:20:20.5 & --30:00:02.8 & 356.11 &  4.04 & 16--05--2013         & 1.29 0.54 \\ 
  15a & 17:46:37.6 & --23:59:57.3 &   4.34 &  2.37 & 01--06--2013         & 1.17 0.52 \\ 
  15b & 17:51:00.3 & --24:00:02.7 &   4.85 &  1.52 & 01--06--2013         & 1.17 0.52 \\ 
  17a & 17:37:51.6 & --24:59:57.3 &   2.43 &  3.54 & 30--06--2013         & 1.10 0.52 \\ 
  17b & 17:42:16.3 & --25:00:02.7 &   2.96 &  2.70 & 30--06--2013         & 1.10 0.52 \\ 
  18a & 17:46:47.6 & --24:59:57.3 &   3.50 &  1.82 & 15--07--2013         & 1.01 0.48 \\ 
  18b & 17:51:12.3 & --25:00:02.7 &   4.01 &  0.97 & 15--07--2013         & 0.98 0.47 \\ 
  20a & 18:07:57.6 & --24:59:57.3 &   5.90 & --2.33 & 01--08--2013        & 1.45 0.62 \\ 
  20b & 18:12:22.3 & --25:00:02.7 &   6.38 & --3.20 & 01--08--2013        & 1.42 0.61 \\ 
  21a & 18:16:54.6 & --24:59:57.3 &   6.87 & --4.11 & 01--07--2013        & 1.54 0.71 \\ 
  21b & 18:21:19.3 & --25:00:02.7 &   7.34 & --4.99 & 01--07--2013        & 1.50 0.70 \\ 
  23a & 17:37:56.5 & --25:59:57.3 &   1.60 &  3.00 & 08--08--2013         & 1.07 0.50 \\ 
  23b & 17:42:23.4 & --26:00:02.7 &   2.13 &  2.15 & 08--08--2013         & 1.07 0.50 \\ 
  24a & 17:49:26.5 & --25:59:57.3 &   2.95 &  0.79 & 11--08--2013         & 1.02 0.49 \\ 
  24b & 17:53:53.4 & --26:00:02.7 &   3.46 & --0.07 & 11--08--2013        & 1.00 0.48 \\ 
  29a & 17:40:03.3 & --26:59:57.2 &   1.00 &  2.06 & 10--07--2013         & 1.24 0.53 \\ 
  29b & 17:44:32.6 & --27:00:02.8 &   1.53 &  1.21 & 10--07--2013         & 1.26 0.53 \\ 
  31a & 17:58:05.3 & --26:59:57.2 &   3.07 & --1.38 & 31--07--2013        & 0.96 0.47 \\
  31b & 18:02:34.6 & --27:00:02.8 &   3.57 & --2.25 & 31--07--2013        & 0.91 0.46 \\
  32a & 18:07:16.3 & --26:59:57.2 &   4.08 & --3.16 & 06--08--2013        & 1.15 0.55 \\
  32b & 18:11:45.6 & --27:00:02.8 &   4.56 & --4.04 & 06--08--2013        & 1.15 0.55 \\
\hline
\multicolumn{7}{l}{Semester P92} \\
\hline 
 32a & 07:06:39.0 & --27:40:57.3 & 239.42 & --9.18 & 26--12--2013         & 1.14 0.53 \\       
 32b & 07:11:10.0 & --27:40:52.2 & 239.86 & --8.28 & 26--12--2013         & 1.08 0.52 \\          
 36a & 08:24:25.4 & --27:40:56.9 & 248.08 &  5.69 & 24--01--2014         &  1.32 0.55\\       
 36b & 08:28:56.3 & --27:40:52.6 & 248.65 &  6.51 & 24--01--2014         &  1.33 0.55\\       
 37a & 08:33:37.2 & --27:40:56.8 & 249.26 &  7.36 & 24--01--2014         &  0.72 0.40\\       
 37b & 08:38:08.2 & --27:40:52.7 & 249.86 &  8.16 & 24--01--2014         &  0.73 0.40\\       
 46a & 08:25:23.3 & --29:41:07.2 & 249.85 &  4.72 & 02--03--2014         &  1.03 0.50\\       
 46b & 08:29:59.5 & --29:41:02.8 & 250.43 &  5.54 & 02--03--2014         &  1.04 0.50\\  
 \hline
 \multicolumn{7}{l}{Semester P93} \\
94a  & 18:11:24.0 & --27:59:57.1 &     272.85 &  --28.00  & 10--04--2014 & 1.08 0.49\\
94b  & 18:15:55.8 & --28:00:02.9 &     273.98 &  --28.00  & 10--04--2014 & 1.09 0.50\\
47a  & 17:11:34.4 & --31:59:57.0 &     257.89 &  --32.00  & 24--06--2014 & 0.81 0.44\\
47b  & 17:16:17.4 & --32:00:03.0 &     259.07 &  --32.00  & 24--06--2014 & 0.84 0.44\\
53a  & 17:11:28.0 & --27:59:57.1 &     257.86 &  --28.00  & 25--08--2014 & 1.17 0.56\\
53b  & 17:15:59.8 & --28:00:02.9 &     258.99 &  --28.00  & 25--08--2014 & 1.16 0.55\\
83a  & 17:46:55.5 & --21:59:57.2 &     266.73 &  --22.00  & 20--06--2014 & 1.07 0.50\\
83b  & 17:51:14.3 & --22:00:02.8 &     267.80 &  --22.00  & 20--06--2014 & 1.06 0.49\\
84a  & 17:55:52.5 & --21:59:57.2 &     268.96 &  --22.00  & 23--06--2014 & 0.97 0.53\\
84b  & 18:00:11.3 & --22:00:02.8 &     270.04 &  --22.00  & 23--06--2014 & 0.95 0.52\\
85a  & 18:04:30.5 & --21:59:57.2 &     271.12 &  --22.00  & 24--06--2014 & 1.15 0.52\\
85b  & 18:08:49.3 & --22:00:02.8 &     272.20 &  --22.00  & 24--06--2014 & 1.15 0.54\\
\hline
\end{tabular}
\end{center}
\end{table*}

\setcounter{table}{0}
\begin{table*}
\begin{center}
\caption{Continued ....}
\begin{tabular}{lcccccc}
\hline
Field & RA      &  DEC    & $l$ & $b$ & Date       & seeing ($^{''}$) \\
      & (J2000) & (J2000) &     &     & (dd-mm-yy) & Mean, $\sigma$ \\
\hline
\multicolumn{7}{l}{Semester P93 (Cont)} \\
\hline
86a  & 18:13:07.5 & --21:59:57.2 &     273.28 &  --22.00  & 01--08--2014 & 1.11 0.50\\
86b  & 18:17:26.3 & --22:00:02.8 &     274.35 &  --22.00  & 01--08--2014 & 1.12 0.51\\
90a  & 17:53:32.6 & --22:59:57.2 &     268.38 &  --23.00  & 30--07--2014 & 0.97 0.47\\
90b  & 17:57:53.2 & --23:00:02.8 &     269.47 &  --23.00  & 30--07--2014 & 0.98 0.47\\
91a  & 18:06:01.6 & --22:59:57.2 &     271.50 &  --23.00  & 30--07--2014 & 0.81 0.43\\
91b  & 18:10:22.2 & --23:00:02.8 &     272.59 &  --23.00  & 30--07--2014 & 0.81 0.43\\
51a  & 17:19:17.3 & --26:59:57.1 &     259.82 &  --27.00  & 01--07--2014 & 1.29 0.55\\
51b  & 17:23:46.6 & --27:00:03.0 &     260.94 &  --27.00  & 01--07--2014 & 1.31 0.56\\
97a  & 17:16:50.0 & --30:59:57.0 &     259.20 &  --31.00  & 27--05--2014 & 1.18 0.54\\
97b  & 17:21:29.9 & --31:00:03.0 &     260.37 &  --31.00  & 27--05--2014 & 1.15 0.52\\
99a  & 17:35:27.0 & --30:59:56.9 &     263.86 &  --31.00  & 23--06--2014 & 0.91 0.45\\
99b  & 17:40:06.9 & --31:00:03.1 &     265.02 &  --31.00  & 23--06--2014 & 0.89 0.45\\
100a & 17:44:49.0 & --30:59:56.9 &     266.20 &  --31.00  & 27--07--2014 & 0.88 0.47\\
100b & 17:49:28.9 & --31:00:03.1 &     267.37 &  --31.00  & 27--07--2014 & 0.89 0.48\\
101a & 17:54:17.0 & --30:59:56.9 &     268.57 &  --31.00  & 29--07--2014 & 1.12 0.51\\
101b & 17:58:56.9 & --31:00:03.1 &     269.73 &  --31.00  & 29--07--2014 & 1.01 0.51\\
102a & 18:03:40.0 & --30:59:56.9 &     270.91 &  --31.00  & 31--07--2014 & 1.21 0.61\\
102b & 18:08:19.9 & --31:00:03.1 &     272.08 &  --31.00  & 31--07--2014 & 1.25 0.63\\
48a  & 17:21:08.5 & --31:59:56.9 &     260.28 &  --32.00  & 23--06--2014 & 1.26 0.56\\
48b  & 17:25:51.4 & --32:00:03.1 &     261.46 &  --32.00  & 23--06--2014 & 1.24 0.58\\
103a & 17:30:59.5 & --31:59:56.9 &     262.74 &  --32.00  & 01--08--2014 & 0.98 0.49\\
103b & 17:35:42.4 & --32:00:03.1 &     263.92 &  --32.00  & 01--08--2014 & 0.99 0.49\\
\hline
\multicolumn{7}{l}{Semester P94} \\
\hline
34a  & 07:29:25.3 & --27:40:57.4 & 112.35  & --27.68 &  14--01--2015& 0.95 0.47\\
34b  & 07:33:56.2 & --27:40:52.1 & 113.48  & --27.68 &  14--01--2015& 0.93 0.46\\ 
39a  & 07:49:12.9 & --31:40:46.6 & 117.30  & --31.67 &  17--12--2014& 0.89 0.47\\
39b  & 07:53:54.9 & --31:40:41.4 & 118.47  & --31.67 &  17--12--2014& 0.87 0.44\\
113a & 07:48:14.4 & --30:40:46.6 & 117.06  & --30.67 &  16--02--2015& 0.79 0.43\\
113b & 07:52:53.4 & --30:40:41.4 & 118.22  & --30.67 &  16--02--2015& 0.77 0.41\\
24a  & 07:22:15.0 & --25:40:49.6 & 110.56  & --25.68 &  24--02--2015& 0.83 0.44\\
24b  & 07:26:41.3 & --25:40:44.3 & 111.67  & --25.67 &  24--02--2015& 0.84 0.44\\ 
40a  & 07:58:12.9 & --31:40:46.5 & 119.55  & --31.67 &  27--12--2014& 0.87 0.46\\ 
40b  & 08:02:54.9 & --31:40:41.5 & 120.72  & --31.67 &  27--12--2014& 0.88 0.46\\ 
107a & 07:22:21.1 & --24:30:23.6 & 110.58  & --24.50 &  23--01--2015& 1.17 0.54\\
107b & 07:26:44.8 & --24:30:18.4 & 111.68  & --24.50 &  23--01--2015& 1.19 0.56\\
108a & 07:31:21.1 & --24:30:23.6 & 112.83  & --24.50 &  26--01--2015& 1.14 0.53\\
108b & 07:35:44.8 & --24:30:18.4 & 113.93  & --24.50 &  26--01--2015& 1.16 0.52\\
38a  & 07:18:52.1 & --23:30:46.6 & 109.71  & --23.51 &  13--01--2015& 0.78 0.43\\ 
38b  & 07:23:13.7 & --23:30:41.4 & 110.80  & --23.51 &  13--01--2015& 0.78 0.43\\ 
42a  & 08:16:12.9 & --31:40:46.5 & 124.05  & --31.67 &  24--02--2015& 0.87 0.46\\ 
42b  & 08:20:54.9 & --31:40:41.6 & 125.22  & --31.67 &  24--02--2015& 0.87 0.46\\ 
115a & 08:12:07.6 & --30:40:46.5 & 123.03  & --30.67 &  13--03--2015& 0.96 0.48\\
115b & 08:16:46.6 & --30:40:41.6 & 124.19  & --30.67 &  13--03--2015& 0.95 0.47\\
116a & 08:21:21.7 & --30:40:46.3 & 125.34  & --30.67 &  22--01--2015& 1.09 0.54\\
116b & 08:26:00.7 & --30:40:41.7 & 126.50  & --30.67 &  22--01--2015& 1.06 0.52\\
25a  & 07:31:12.5 & --25:40:49.5 & 112.80  & --25.68 &  22--03--2015& 1.05 0.50\\ 
25b  & 07:35:38.7 & --25:40:44.1 & 113.91  & --25.67 &  22--03--2015& 1.04 0.50\\ 
33a  & 07:36:37.5 & --23:30:46.5 & 114.15  & --23.51 &  13--01--2015& 0.86 0.45\\ 
33b  & 07:40:59.1 & --23:30:41.5 & 115.24  & --23.51 &  13--01--2015& 0.87 0.45\\ 
36a  & 08:07:12.9 & --31:40:46.5 & 121.80  & --31.67 &  23--02--2015& 0.90 0.45\\ 
36b  & 08:11:54.9 & --31:40:41.5 & 122.97  & --31.67 &  23--02--2015& 0.92 0.46\\ 
112a & 07:37:56.4 & --30:40:46.6 & 114.48  & --30.67 &  14--01--2015& 0.89 0.45\\
112b & 07:42:35.4 & --30:40:41.4 & 115.64  & --30.67 &  14--01--2015& 0.88 0.45\\
117a & 07:39:54.9 & --31:40:46.7 & 114.97  & --31.67 &  24--01--2015& 0.99 0.47\\
117b & 07:44:36.9 & --31:40:41.3 & 116.15  & --31.67 &  24--01--2015& 1.00 0.47\\
109a & 07:39:56.5 & --24:30:46.5 & 114.98  & --24.51 &  23--02--2015& 0.92 0.46\\
109b & 07:44:20.2 & --24:30:41.5 & 116.08  & --24.51 &  23--02--2015& 0.94 0.47\\
114a & 08:02:50.4 & --30:40:46.5 & 120.71  & --30.67 &  25--02--2015& 0.99 0.49\\
114b & 08:07:29.4 & --30:40:41.5 & 121.87  & --30.67 &  25--02--2015& 1.00 0.50\\
\hline
\end{tabular}
\end{center}
\end{table*}

\begin{figure*}
\begin{center}
\setlength{\unitlength}{1cm}
\begin{picture}(18,22)
\put(-3,-2){\includegraphics{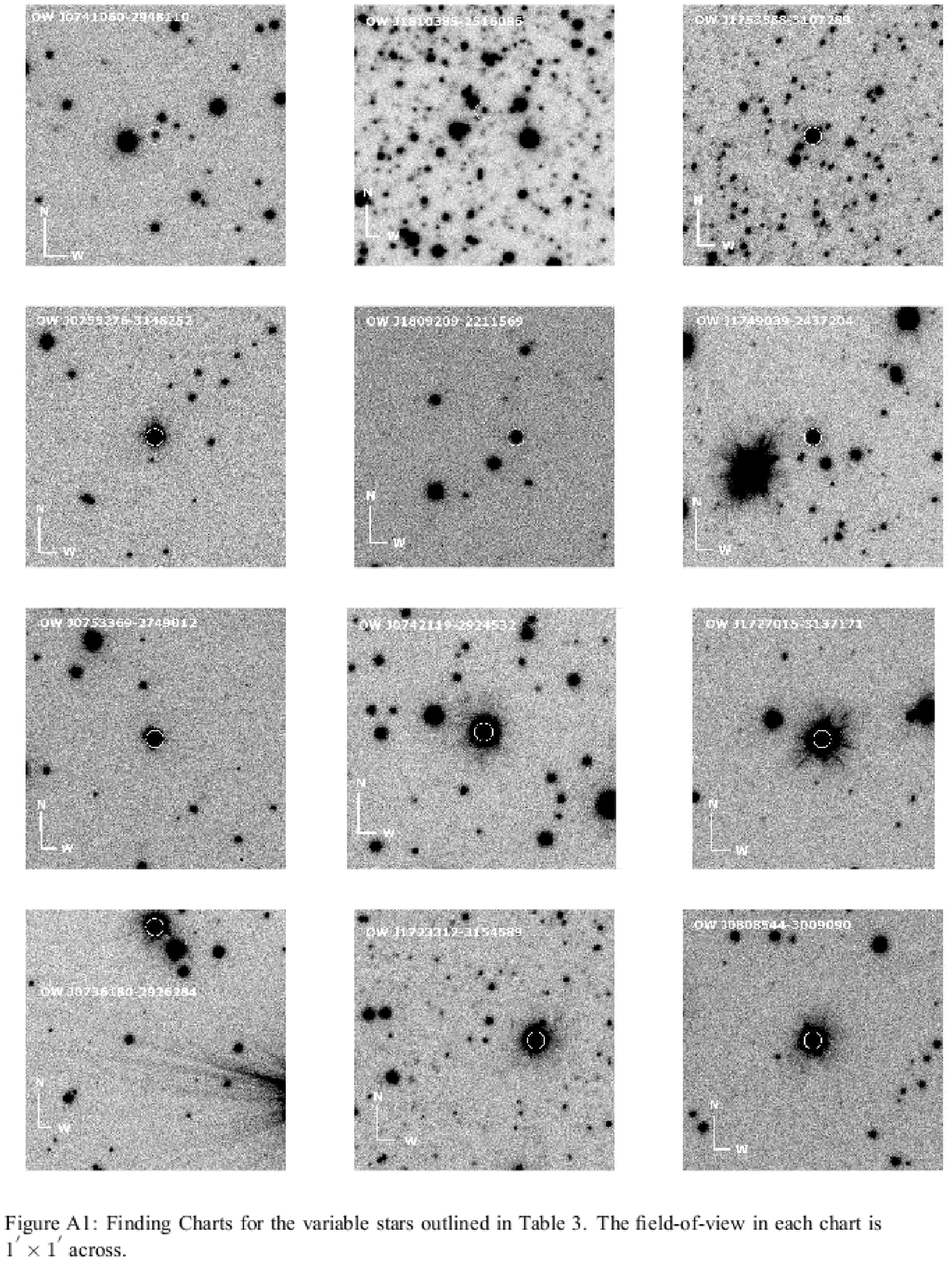}}
\end{picture}
\end{center}
\label{finding} 
\end{figure*}

\setcounter{figure}{0}
\begin{figure*}
\begin{center}
\setlength{\unitlength}{1cm}
\begin{picture}(18,17)
\put(-3,-6){\includegraphics{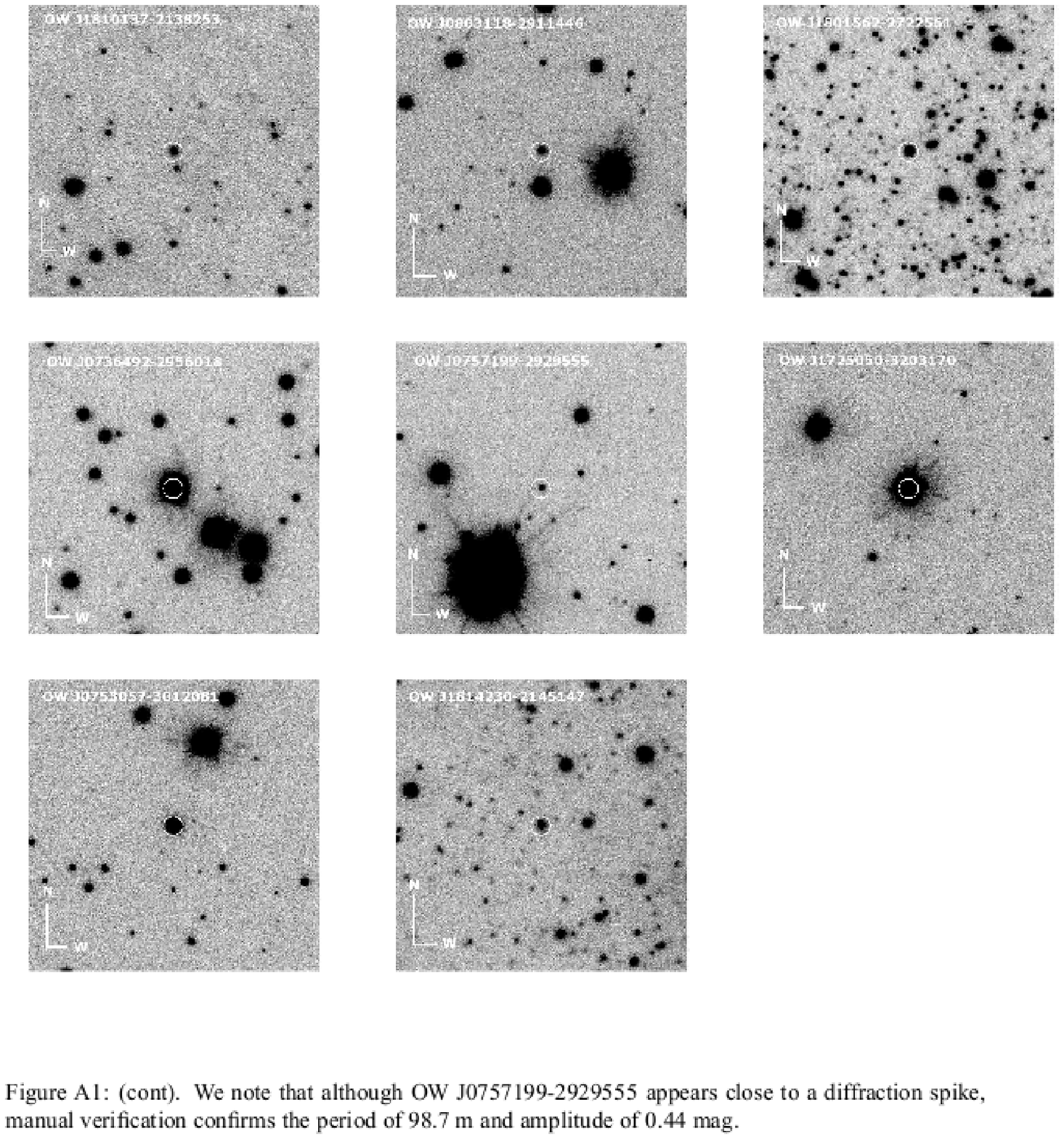}}
\end{picture}
\end{center}
\label{finding} 
\end{figure*}

\end{document}